\documentclass[11pt]{article}
\usepackage[fleqn]{amsmath}
\usepackage{amsthm,amssymb}
\usepackage{graphicx}
\usepackage{hhline}
\usepackage{cite}

\makeatletter
\@addtoreset{equation}{section}
\makeatother

\def\espneg{\!\!\!\!\!}

{\theoremstyle{remark}
}

\newcommand{\bra}[1]{\langle\,#1\,|}
\newcommand{\ket}[1]{|\,#1\,\rangle}
\newcommand{\bracket}[1]{\langle\,#1\,\rangle}

\def\vac{\mathrm{vac}}

\def\le{\leqslant}

\def\sul{\sum\limits}
\def\pl{\prod\limits}

\def\build#1_#2^#3{\mathrel{
\mathop{\kern 0pt#1}\limits_{#2}^{#3}}}

\newcommand{\Zset }{{\mathbb Z}}

\def\cR{\mathcal{R}}

\def\cC{\mathcal{C}}

\def\cO{\mathcal{O}}

\usepackage{euscript}

\renewcommand{\H}{{\boldsymbol{\mathrm{H}}}}
\def\J{{\boldsymbol{\mathnormal{J}}}}

\def\eps{\varepsilon}

\def\sg{\sigma}

\def\vth{\vartheta}
\def\sg{\sigma}

\def\tend{\rightarrow}

\begin{document}

\begin{titlepage}
\begin{flushright}
RUNHETC-2002-18\\
\end{flushright}

\vspace{2cm}

\begin{center}
\begin{LARGE}
{\bf Long-distance asymptotics of spin-spin correlation functions
for the XXZ spin  chain}
\end{LARGE}

\vspace{1.5cm}

\begin{large}

{\bf Sergei Lukyanov}$^{a,b}$ {\bf and}
{\bf V\'eronique Terras}$^{a,c}$
\end{large}

\vspace{.5cm}
 
{$^{a}$NHETC, Department of Physics and Astronomy\\
     Rutgers University\\ 
     Piscataway, NJ 08854, USA\\
\vspace{.5cm}
 $^{b}$L.D. Landau Institute for Theoretical Physics\\
     Kosygina 2, Moscow, Russia\\
\vspace{.5cm}
 $^{c}$LPMT, UMR 5825 du CNRS\\
     Universit\'e Montpellier II,
     Montpellier, France}
\vspace{2cm}

\centerline{\bf Abstract} \vspace{.8cm}
\parbox{12cm}{We study asymptotic expansions of 
spin-spin correlation functions for the XXZ Heisenberg chain in the 
critical regime. We use the fact that
the  long-distance effects can be described by the Gaussian conformal
field theory.
Comparing exact results for form factors in the XYZ and sine-Gordon 
models, we  determine correlation amplitudes for the leading and main 
sub-leading terms in the asymptotic expansions of spin-spin correlation
functions. We also study the isotropic (XXX) limit of these expansions.
}
\end{center}

\vfill
 
 
\end{titlepage}
\newpage

\section{Introduction}

In the domain of two-dimensional  integrable models, 
it is, in general, still a challenging
problem to compute 
correlation functions in the form of compact and manageable expressions.
For lattice systems, 
a few methods of computation have been developed 
\cite{BogIK93L,JimM95L,KMT,KMST}: in particular, it is possible
in some cases to obtain exact integral representations 
of correlation functions.
However it is still quite difficult to analyse those
expressions, and especially to extract their long-distance asymptotic 
behavior.

On the other hand, at the critical point,
when the gap in the spectrum of 
the lattice Hamiltonian $ \H_{\text{latt}}$ vanishes, 
the correlation length becomes
infinite in units of the lattice spacing. As a result, 
the leading long-distance effects  can be described by 
Conformal Field Theory (CFT) with the  Hamiltonian $\H_{\text{CFT}}$
\cite{Polyakov}.
In general, it is expected
that  the critical  lattice Hamiltonian admits
an asymptotic power series expansion
involving an infinite set of local scaling fields\ (see e.g.
\cite{Pokrov,Kadanoff}):
\begin{equation}\label{densDVT}
\H_{\text{latt}}\sim \H_{\text{CFT}}+\lambda_{1} \, \eps^{d_{1}-2} \int dx\,
\cO_{1} +  \lambda_{2} \, \eps^{{d_2}-2}\int dx\,
\cO_{2}    + \dots\ .
\end{equation}
Here $d_{k}$ denotes the scaling dimension of the field
$\cO_{k}$, and
the explicit dependence on the lattice spacing 
$\varepsilon$ is used to show the relative smallness of various terms.
All  fields occurring in
\eqref{densDVT} are irrelevant  and 
the corresponding exponents of $\eps$
are positive.
Such an  expansion also  involves a set of non-universal
coupling constants $\lambda_{k}$ depending on the microscopic properties
of the model, and whose values rely on the chosen normalization for
$\cO_{k}$.

The asymptotic expansion~\eqref{densDVT} is a powerful
tool to study the low energy spectrum of  lattice theories.
At the same time,  in order to analyse
lattice correlation functions, one should also know how
local lattice operators $\cO_{\text{latt}}$ are 
represented in terms of scaling fields.
Just as the Hamiltonian density, they
can be expressed as formal series in powers of the lattice spacing $\eps$, 
\begin{equation}\label{corrDVT}
   \cO_{\text{latt}}\sim C_{m_1}\, \eps^{d_{m_1}} \cO_{m_1}
                   + C_{m_2}\, \eps^{d_{m_2}} \cO_{m_2}    + \dots\ ,
\end{equation}
where again the non-universal  constants $C_m$  depend on the 
normalization of the scaling fields.

The knowledge of the explicit form of \eqref{densDVT} and \eqref{corrDVT}
enables one to obtain the long-distance asymptotic expansion of
lattice correlation functions in   powers 
of  lattice distances.
For example, in the case of a vacuum  correlator,
the leading term 
is given by the CFT  correlation function of the
operator occurring in \eqref{corrDVT} at the lowest order in $\eps$,
whereas subleading asymptotics come from both higher order terms 
in \eqref{corrDVT} and from the perturbative
corrections
to the CFT vacuum toward the lattice ground state.
In this expansion, the exact values of the exponents follow
from the scaling dimensions of the CFT fields, 
while the corresponding  amplitudes come from the
values of the constants appearing in  \eqref{densDVT}
and  \eqref{corrDVT}.
Thus, to investigate quantitatively the long-distance behavior of lattice 
correlation functions,
one needs 
to determine  (for a fixed  suitable normalization of the scaling fields) 
exact analytical 
expressions for the non-universal constants $\lambda_{k}$
and $C_{m}$.

The aim of this article is to study expansions of the type\ \eqref{corrDVT}
in the case of 
the XXZ spin 1/2 Heisenberg chain.
In particular, we calculate the first constants
$C_{m}$ occurring in the expansion \eqref{corrDVT} of local spin operators.

\bigskip

The article is organized as follows.
 
In Section~\ref{secXXZ}, we recall the definition
of the XXZ spin  chain\ (see e.g. \cite{Bax82L} for details), 
and  discuss
its continuous limit, the Gaussian CFT\ \cite{LutP75,KadBr,Aff89}.
We review in particular how, from the analysis of the 
global symmetries of the model, 
one can derive selection rules for the expansions \eqref{corrDVT} 
of the lattice operators in terms of the scaling fields of the 
Gaussian model. This enables one to predict the  structure of the 
long-distance asymptotic expansions of the spin-spin correlation functions.
 
The problem of computing the constants occurring 
in the expansions \eqref{corrDVT} of spin operators is the subject of 
Section~\ref{secDVT}. There we explain how, moving slightly away from 
criticality, and comparing  exact results  obtained
for the XYZ and sine-Gordon models,
one can quantitatively connect lattice spin operators to scaling
fields. 
This gives us access 
to
the correlation amplitudes 
of the spin-spin correlation functions.
Our predictions are gathered in Section~\ref{secASY}, where they are
compared to existing numerical data.

In Section~\ref{secXXX}, we study   these spin-spin correlation
functions in the isotropic (XXX) limit by means of the exact 
Renormalization Group (RG) approach. 
This section contains an erratum of Section 5 
of \cite{Luk98}.

Finally, we conclude this article with several remarks.

\section{XXZ spin  chain in the continuous limit}
\label{secXXZ}

\subsection{Preliminaries\cite{Bax82L,Aff89}}

To illustrate the problem of the
determination of correlation amplitudes,
we consider an example,
the XXZ Heisenberg chain of spins 1/2.
It is defined by the Hamiltonian
\begin{equation}\label{Ham}
    \H_{\mathrm{XXZ}}= -\frac{J}{2} \sul_{l=-\infty}^{\infty}
            \Bigl\{ \sg_l^x \sg_{l+1}^x + \sg_l^y \sg_{l+1}^y +
                            \Delta\, \sg_l^z \sg_{l+1}^z \Bigr\}\, ,
\end{equation}
where the spin operators $\sg_l^a$ ($a=x,y,z$) denote
the conventional Pauli matrices associated with the $l$-th site of the
infinite lattice,
and $\Delta$ is an  anisotropy parameter.
As $\H_{\mathrm{XXZ}}(-J,-\Delta)$ can be obtained from
$\H_{\mathrm{XXZ}}(J,\Delta)$ by a unitary transformation,
we\ choose in the following the coupling constant $J$ to be positive.

The nature of the spectrum of the infinite chain depends on
the value of the anisotropy parameter $\Delta$.
The study we present here concerns the critical regime of
the chain, which corresponds to the domain $-1\le \Delta  <1$.
We shall use the parameterizations
\begin{equation}\label{Delta}
      \Delta=  \cos (\pi \eta)\ \qquad\ ( 0<\eta \leq 1)\, ,
\end{equation}
and
\begin{equation}\label{constJ}
   J=\frac{1-\eta}{\eps \sin(\pi\eta)} \qquad (\eps >0)\, .
\end{equation}

To describe  leading long-distance (low-energy)
effects,
it is useful to consider a continuous limit of the lattice model.
This limit can  be obtained from the representation of \eqref{Ham}
in terms of  lattice  fermions,
\begin{equation}\label{Hferm}
   \H_{\mathrm{XXZ}}= -J\ \sum_l
            \Bigl\{\, \psi_l^{\dagger} \psi_{l+1}^{}
                  + \psi_{l+1}^{\dagger} \psi_{l}^{}
                  + \frac{\Delta}{2}\
                       \big(1-2\psi^{\dagger}_l\psi_l^{}\big)\,
    \big(1-2\psi^{\dagger}_{l+1}
\psi_{l+1}^{}\big)\, \Bigr\}\, ,
\end{equation}
where the fermionic operators are
related to the spin operators through the  Jordan-Wigner
transformation,
\begin{equation}
  \sg_{l}^{z}=1-2\psi_{l}^{\dagger}\psi_{l}^{}\, ,\qquad
  \psi_{l}^{\dagger}= 
\prod_{j<l}\sigma^z_j\cdot  \sg_{l}^{-}\, ,\qquad
  \psi_{l}^{}    =\prod_{j<l}\sigma^z_j\cdot  
 \sg_{l}^{+}\, ,
\end{equation}
and  $\sg_{l}^{\pm}=(\sg_l^x\pm i\sg_l^y)/2\, .$
With the parameterization \eqref{constJ} of 
the  constant $J$,
the spin-1/2 spin  wave
dispersion relation has the form\ \cite{JohKM73,Fadeev} 
\begin{equation}\label{disp}
      {\cal E}(k)= -\frac{\cos(k)}{\varepsilon}\, .
\end{equation}
The ground state of the chain has all levels filled with  $|k|<\pi/2$.
Linearizing the  dispersion relation in the vicinity of the
Fermi points $k_F=\pm\pi/2$,
one can take the continuous limit of the model in which
the  lattice operators $\psi_{l}$ are replaced by two
fields, $\psi_{R} (x)$ and $\psi_{L} (x)$, varying slowly
on the lattice scale:
\begin{equation}\label{fermRL}
   \psi_{l}\propto  e^{\frac{i\pi l}{2}}\,
                   \psi_{R}(x) + e^{-\frac{i\pi l}{2}}\,
\psi_{L}(x)\, .
\end{equation}
Here $x=l\, \varepsilon$, and thus
the parameter $\eps$ can  be interpreted as a
lattice spacing.
The continuous
Fermi fields are governed by the  Hamiltonian
\begin{equation}\label{Thir}
  {\bf H}_{\mathrm{ Thirring}}=
       \int \frac{dx}{2\pi} \,   \Bigl\{
            -i\psi^{\dagger}_R \partial_x\psi_R^{}
            +i\psi^{\dagger}_L \partial_x\psi_L^{}
      +{\tt g}\, \psi^{\dagger}_R\psi_R^{} \psi^{\dagger}_L\psi_L^{}\,
                                  \Bigr\}\, .
\end{equation}
A precise relation between the anisotropy parameter $\Delta$ and
the four-fermion coupling constant ${\tt g}$ depends on the choice of the
regularization procedure for the continuous Hamiltonian\ \eqref{Thir}\
and is not essential for
our purposes. What is important is that the quantum field theory 
model\ \eqref{Thir}
(which is known as the Thirring model\ \cite{Thirring,Klaiber}) is 
conformally invariant
and equivalent to the Gaussian CFT\ (see
the preprint collection\cite{Coleman} for a historical 
review of bosonization).

\subsection{Exponential fields in the Gaussian model}

The Gaussian  model  is   defined in
terms of one   scalar field $\varphi$,
which satisfies
the D'Alembert equation\footnote{
The value of the  spin-wave velocity $v$ is determined from the slope of the
dispersion relation\ \eqref{disp}\ at the Fermi points  and
from the identification of
$x$ with $l\, \varepsilon$.}
\begin{equation}\label{vsfdr}
      (\partial_t^2-v^2\partial_x^2)\, \varphi=0\, ,
                 \qquad \text{with}\ \ v=1\, ,
\end{equation}
and the  boundary condition
\begin{equation}\label{bound}
\partial_t\varphi(x,t)|_{x\to\pm \infty}= 
\partial_x\varphi(x,t)|_{x\to\pm \infty}=0\ .
\end{equation}
Assuming that the  equal-time
canonical commutation relations are imposed on the field,
\begin{equation*}
        [\, \varphi(x) ,\partial_t\varphi(x')\, ]=8\pi i\ \delta(x-x')\, ,\
\end{equation*}
one can write the Hamiltonian 
of the  model in the form
\begin{equation}
   \H_{\mathrm{Gauss}}=
         \int \frac{dx}{2\pi} \Big\{ \mathbf{T}_R + \mathbf{T}_L
        \Big\}+\text{const}\, ,\label{Hgauss}
\end{equation}
where
\begin{equation}
\mathbf{T}_R(x)=\frac{1}{16}  (\partial_x\varphi-\partial_t\varphi)^2\, ,
   \ \qquad\ 
   \mathbf{T}_L(x)=\frac{1}{16}  (\partial_x\varphi+\partial_t\varphi)^2\, .
\end{equation}

As is usual in CFT\ \cite{Belavin},
it is convenient to set a class of conformal
primary fields
among  all  scaling fields. In the case  of the Gaussian model,
these  conformal primaries  include
right and left currents,
\begin{equation}\label{dPhi}
        (\partial_x-\partial_t)\varphi\, , \ \qquad\ 
        (\partial_x+\partial_t)\varphi\, ,
\end{equation}
along with   exponential fields\footnote{Here we are not considering 
the orbifold (Ashkin-Teller) sector of the Gaussian model.} \cite{KadBr}. 
With  the
boundary condition\ \eqref{bound}, the latter can  be defined
as
\begin{equation}\label{op}
   \cO_{s,n}(x,t)=\Lambda^{d_{s,n}} \lim_{\varepsilon\to +0}\,
     \exp\bigg\{\frac{i n}{4\sqrt{\eta}}
                \int_{-\infty}^x dx'\, \partial_t\varphi(x',t)\bigg\}\,
     \exp\bigg\{ \frac{is\sqrt{\eta}}{2}\ \varphi(x+\eps,t) \bigg\}\, ,
\end{equation}
where $s$, $n$ are integers and
\begin{equation}\label{dimen}
       d_{s,n}= \frac{s^2\eta}{2}+\frac{n^2}{8\eta}\ .
\end{equation}
The regularization
parameter $\Lambda$, which has the dimension $[\, \text{length}\,]^{-1}$,
is introduced
in the definition\ \eqref{op}\
in order to provide a multiplicative renormalization of the fields.
Notice that 
$\cO_{s,n}$ obey the simple Hermiticity relation
\begin{equation}\label{hermit}
         \cO_{s,n}^{\dagger}=\cO_{-s,-n}\ .
\end{equation}
To completely define these exponential fields,
we should also specify some  condition which fixes their
multiplicative normalization.
By a proper choice of $\Lambda$ in \eqref{op}, 
one can impose the following form of the causal  Green's functions
in the Euclidean domain $x^2-t^2>0$:
\begin{equation}\label{CFTnorm}
   \bracket{ T\, \cO_{s,n}(x,t)\, \cO^{\dagger}_{s,n}(0,0)}
= \Big(\frac{t-x}{t+x}\Big)^{\frac{sn}{2}}\ (x^2-t^2)^{-d_{s,n}}\, .
\end{equation}
We will later refer to Eq.~\eqref{CFTnorm} as the 
``CFT normalization condition''.

\subsection{Global  symmetries}

To draw a precise link between the Gaussian CFT and the XXZ spin chain, 
it is important
to examine and identify their global symmetries.

Let us first consider the Gaussian model.
The Hamiltonian\ \eqref{Hgauss}  is
manifestly invariant under the $U(1)$ rotations,
\begin{equation}\label{trU}
 {\mathbb U}_{\alpha}\, \varphi\,
                     {\mathbb U}^{-1}_{\alpha}
     =\varphi+\frac{2\alpha}{\sqrt\eta}\ ,  \qquad\quad 
 {\mathbb U}_{\alpha}\, \cO_{s,n}\, {\mathbb U}^{-1}_{\alpha}=
                     e^{is\alpha}\,  \cO_{s,n}\ ,
\end{equation}
where   the operator ${\mathbb U}_{\alpha}$ can be written in the form
\begin{equation}\label{SZZA}
   {\mathbb U}_{\alpha}
      =\exp\bigg\{\frac{i\alpha}{4\pi\sqrt{\eta}}\
                \int_{-\infty}^{\infty}dx\, \partial_{t}\varphi\, \bigg\}\ .
\end{equation}
The CFT model is also  invariant under 
the parity transformation
${\mathbb P}\, :\,\varphi(x,t)\to \varphi(-x,t),$
the time reversal ${\mathbb T}\, :\, \varphi(x,t)\to \varphi(x,-t), $
and the  reflection ${\mathbb C}\, :\, \varphi\to -\varphi$.
Using the definition\ \eqref{op}, one can show 
that these transformations act on the
exponential fields as
\begin{align}\label{trP}
  &{\mathbb P}\, \cO_{s,n}(x,t)\, {\mathbb P}
 =e^{-\frac{is n\pi}{2}}\ {\mathbb U}_{\pi n}\ \cO_{s,-n}(-x,t)\, ,\\
  \label{trT}
  &{\mathbb T}\, \cO_{s,n}(x,t)\, {\mathbb T}= \cO_{-s,n}(x,-t)\, ,\\
  \label{trC}
  &{\mathbb C}\, \cO_{s,n}\, {\mathbb C}= \cO_{-s,-n}(x,t)\, .
\end{align}
Note that the  reflection ${\mathbb C}$ can be naturally considered
as a charge conjugation in the theory. As usual, 
${\mathbb C}$ and ${\mathbb P}$ are  intrinsic automorphisms
of the operator algebra, contrary to the anti-unitary
transformation  ${\mathbb T}$ which acts on  $c$-numbers as follows,
\begin{equation*}
   {\mathbb T}\, (\text{$c$-number})\, {\mathbb T}
   =(\text{$c$-number})^*\ .
\end{equation*}

\bigskip

Let us now identify
the above transformations
with the global  symmetries of the spin chain.
First of all, it is natural to define
the action of the $U(1)$ rotation
on the lattice as follows,
\begin{equation}\label{Ulatt}
     {\mathbb U}_{\alpha}\, \sigma^{\pm}_l\,{\mathbb U}^{-1}_{\alpha}=
                   e^{\pm i\alpha}\ \sigma^{\pm}_l\, ,   \qquad\quad
     {\mathbb U}_{\alpha}\, \sigma^{z}_l\, {\mathbb U}^{-1}_{\alpha}=
                  \sigma^{z}_l\ ,
\end{equation}
where
\begin{equation}\label{ups}
   {\mathbb  U}_{\alpha}=e^{i\alpha S_z}\, \
      \qquad\qquad\ \ \text{with}\quad 
   S_z=\frac{1}{2}\,  \sum_{j}\sigma^z_j\ .
\end{equation}
Such an  identification has an important consequence.
Indeed, as the two rotations of angle $2\pi$ and $-2\pi$ are 
indistinguishable
trivial transformations in the lattice theory, we should set
\begin{equation}\label{sign}
     {\mathbb U}_{2\pi}= {\mathbb U}_{-2\pi}=\nu\, {\mathbb I}\,  
 \qquad {\rm with}
 \quad \nu=\pm 1
\end{equation}
in the corresponding Gaussian model.
The condition\ \eqref{sign}, along with\ \eqref{trU},
implies that    $\varphi$ has to be
treated as a compactified field:
\begin{equation*}
   \varphi\equiv \varphi +\frac{4\pi}{\sqrt{\eta}}\  {\mathbb Z}\ .
\end{equation*}
According to \eqref{ups}, the  sign factor $\nu=+1$ in Eq.~\eqref{sign} 
occurs for states with an  integer  eigenvalue of the
operator $S_z$. The corresponding  
linear subspace  of the whole Hilbert space 
can be constructed from the thermodynamic limit
of finite chains with an even number of sites, 
and will therefore be referred to 
as the ``even sector''. Similarly,  the condition  $\nu=-1$
defines  another linear subspace  which will be called the  ``odd sector''.

The actions of charge conjugation and time reversal
can be naturally identified,
in the lattice theory, with the following transformations,
\begin{alignat}{2}\label{trPlat}
        &{\mathbb C}\, \sigma^{\pm}_l\, {\mathbb C}=\sigma^{\mp}_{l}\, ,
           \qquad\qquad  
        &&{\mathbb C}\, \sigma^{z}_l\, {\mathbb C}=-\sigma^{z}_{l}\, ,\\
&{\mathbb T}\, \sigma^{\pm}_l\, {\mathbb T} =
\sigma^{\mp}_l\, , &&
{\mathbb T}\, \sigma^{z}_l \, {\mathbb T} =  -\sigma^{z}_l\ . 
\end{alignat}
Let us recall here 
that the  time reversal is an anti-unitary transformation, so
that ${\mathbb C}$ and ${\mathbb T}$ correspond to different 
symmetries of the XXZ chain,  even thought
they act  identically  on the spin operators.

As for the parity transformation, its action in the lattice model  depends
on the choice of the sector specified by the sign
factor $\nu$ in\ \eqref{sign}. Indeed,
$\nu=+1$ implies that the considered
infinite chain is defined as the thermodynamic limit of 
finite lattices with an even number of sites. 
Such finite lattices clearly do not possess 
any invariant  site  with respect to the parity transformation.
Therefore, 
\begin{equation}\label{gare}
        {\mathbb P}\, \sigma^a_l\, {\mathbb P}=\sigma^a_{1-l}\
                  \qquad\ {\rm for}\qquad  \nu=+1\, ,
\end{equation}
whereas the ``naive''  action of ${\mathbb P}$
is  valid only in the  odd sector:
\begin{equation}\label{garek}
         {\mathbb P}\, \sigma^a_l\, {\mathbb P}=\sigma^a_{-l}\
                  \qquad\  \ {\rm for}\qquad  \nu=-1\, .
\end{equation}


\subsection{Selection rules}

The global  symmetries and the knowledge of their action 
on  lattice and  continuous operators
provide selection rules for the set of
scaling fields which can occur in expansions \eqref{densDVT} and
\eqref{corrDVT}.
Here we examine what are these
selection rules for the conformal primary
fields in the expansions  of the  spin operators.

First, since the whole operator content of
the  Gaussian model is given by
the  primary fields described above and by their conformal descendants,
it is easy to see that $\cO_{\pm 1,0}$ are the local fields
with the lowest scaling dimension which
may occur in the expansions of  $\sigma_0^{\pm}$:
\begin{equation}\label{expanv}
   \sigma_0^{\pm}\sim C_0\ \varepsilon^{d_{1,0}}\ \cO_{\pm 1,0}(0)+\ldots\ .
\end{equation}
Furthermore, assuming that the conjugation in the lattice theory is defined
in such a way that $\sigma_{l}^{a}\ (a=x,y,z)$ are Hermitian operators,
we conclude
from \eqref{hermit} that the constant  $ C_0$ in\ \eqref{expanv}\
is real.

Let us now consider the  expansion of the lattice operator $\sigma_0^{z}$.
Due to the $U(1)$ invariance, it can contain
only the primary  fields $(\partial_x\pm \partial_t)\varphi$ and
$\cO_{0,n}$, $n\in\Zset$, along with their conformal descendants.
Using definition\ \eqref{op}, it is
easy to show that  the fields $\cO_{0,n}$ with  odd
$n$ are not  mutually local
with respect to $\cO_{\pm 1,0}$. As the latter are  the leading terms
in the expansions\ \eqref{expanv} of $\sigma_0^{\pm}$,
it would  contradict
the mutual locality of the lattice operators $\sigma_j^{\pm}$ and
$\sigma_{l}^{z}$ if there were any $\cO_{0,2m+1}$, $m\in\Zset$, 
in the series for $\sigma_0^{z}$. 
{}From the  $ {\mathbb C},\ {\mathbb T}$ invariances and
the Hermiticity of $\sg_0^{z}$,  one can moreover
predict that the primary fields are allowed to appear
only as linear  combinations of
$\partial_t\varphi$   and
$i\, (\cO_{0,2m}-\cO_{0,-2m})$ with real coefficients.
Finally, let us consider the parity transformation.
According to equations\ \eqref{trP} and \eqref{sign},
\begin{equation}\label{hyt}
   {\mathbb P}\, \cO_{s,2m}(x,t)\, {\mathbb P}
                 =\nu^{m}\ \cO_{s,-2m}(-x,t)\, ,
\end{equation}
and thus one has to examine even and odd sectors separately.
In the  odd sector ($\nu=-1$), 
the parity\  \eqref{garek} prohibits the
presence of the fields $\cO_{0,4m}-\cO_{0,-4m}$ with $m\in\Zset$.
Therefore the expansion has to be of the form,
\begin{align}\label{uytra}
        \sg_0^{z}\sim\,  &\,
         \varepsilon\, C_0^z\, \partial_t \varphi(0)   \nonumber \\
        &+\frac{1}{2 i}\,\sul_{m=1}^{+\infty}\,  C^z_{m}\
            \eps^{d_{0,4m-2}}\  \big(\cO_{0,4m-2}-\cO_{0,-4m+2}\big)(0)
            +\, \text{descendants}\, .
\end{align}
But, since \eqref{uytra} is a  local operator  expansion, 
it cannot depend on the
choice of the sector in the Hilbert space. 
Thus,  \eqref{uytra} should be   valid in the even sector as well.
Owing to this and to Eqs. \eqref{gare}, \eqref{hyt}, we can  determine
the action of the  lattice translation,
\begin{equation}
   \mathbb{K}\, \sg_l^a\, \mathbb{K}^{-1}= \sg_{l+1}^a\, ,
\end{equation}
on the primary  fields $\cO_{0,4m-2}$:
\begin{equation}\label{vdctt}
   \mathbb{K}\, \cO_{0,4m-2}(x)\,
                \mathbb{K}^{-1}=- \cO_{0,4m-2}(x+\varepsilon)\, .
\end{equation}
With  this relation, the expansion  \eqref{uytra}\  can
be written in  
the more general form,
\begin{align}\label{dvtz}
   \sg_l^{z}\sim &\,
         \frac{\varepsilon}{2\pi\sqrt{\eta}}\,
 \partial_t \varphi(x)  \nonumber  \\
    & +\frac{(-1)^{l}}{2 i}\,
  \sul_{m=1}^{+\infty}\,  C^z_{m}\
        \eps^{d_{0,4m-2}}\,  \big(\cO_{0,4m-2}-\cO_{0,-4m+2}\big)(x)
            +\, \text{descendants}\, ,
\end{align}
where $x=l\, \eps$. Notice that the exact value of the coefficient
$C_0^z$ in  \eqref{dvtz} is  determined
from the comparison of Eqs.~\eqref{SZZA}
and\  \eqref{ups}.

Using the same line of arguments, one can extend the 
expansion\ \eqref{expanv},
\begin{equation}\label{dvt+-}
  \sg_l^{\pm} \sim\, 
  \frac{1}{2}\, \sul_{m=0}^{+\infty} (-1)^{lm}\,  C_m\
          \eps^{d_{1,2m}}\ \big(\cO_{\pm 1,2m}+\cO_{\pm 1,-2m}\big)(x)
            +\ \text{descendants}\ ,
\end{equation}
and determine the action of the lattice translation
$ \mathbb{K}$ on $\cO_{\pm 1,2m}$:
\begin{equation}\label{nshyt}
   \mathbb{K}\, \cO_{\pm 1 ,2m}(x)\,
\mathbb{K}^{-1}=(-1)^m\ \cO_{\pm 1 ,2m }(x+\varepsilon)\ .
\end{equation}
Since all exponential fields $\cO_{\pm s ,2m}(x)$ with 
integers $s$ and $m$
can be obtained by means of  operator product expansions
of the fields $\cO_{ \pm 1 ,2m}(x)$
and  $\cO_{0 ,4m-2}(x)$, we deduce 
from\ \eqref{vdctt}\ and\ \eqref{nshyt}
that
\begin{equation}\label{newsd}
   \mathbb{K}\, \cO_{ s ,2m}(x)\,
\mathbb{K}^{-1}=(-1)^m\ \cO_{s ,2m }(x+\varepsilon)\ .
\end{equation}
Notice that, in the process of bosonization,
the fermionic fields $\psi_R$ and $\psi_L$ \eqref{fermRL}
are  identified with the
exponential fields $\cO_{1,-1}$ and $\cO_{1,1}$ respectively.
Hence equation \eqref{newsd}\ is indeed 
consistent with\ \eqref{fermRL}.

\bigskip

It is not difficult to extend our symmetry analysis
to derive  the selection rules for
expansion\ \eqref{corrDVT}\ of any local lattice operator.
This procedure can, in particular, be applied to the case of the 
Hamiltonian  density. As a result,
the following form
of the low energy effective Hamiltonian for the XXZ spin chain is suggested:
\begin{equation}\label{nhst}
     \H_{\mathrm{XXZ}}\sim
         \H_{\text{Gauss}}+\int \frac{dx}{4\pi}
             \bigg\{\, \sum_{m=1}^{\infty}
         \lambda_m\, \varepsilon^{d_{0,4m}-2}\,
   \big(\cO_{0,4m}+\cO_{ 0,-4m}\big)(x)+{\rm descendants}\, \bigg\}\, .
\end{equation}

As was mentioned in Introduction, it is necessary
to choose the normalization of the scaling fields to give a precise
meaning to the couplings   in \eqref{nhst} as well as to
the real constants $C_m,\ C_m^z$ in expansions
\eqref{dvtz} and \eqref{dvt+-}. In this paper we
adapt the  CFT normalization\ \eqref{CFTnorm}.
With this normalization condition, the first coupling  constant $\lambda_1$
in\ \eqref{nhst}\ has been obtained in \cite{Luk98},
together with
the leading contributions of descendant fields
(see also Ref.\cite{Cardy} for a qualitative 
analysis of the  descendent field
contributions):
\begin{multline}\label{Heff}
    \H_{\mathrm{XXZ}}\sim
     \H_{\text{Gauss}}+\lambda_1\ \eps^{2/\eta-2}\, \int \frac{dx}{4\pi}\,
\big(\cO_{0,4}+\cO_{0,-4}\big)(x)\\
        -\eps^2\int \frac{dx}{2\pi}\,
        \Big\{\,  \lambda_+\  \mathbf{T}_R\mathbf{T}_L(x)
        + \lambda_-\  \big(\mathbf{T}^2_R+\mathbf{T}^2_L\big)(x)\, \Big\}
+\dots\, ,
\end{multline}
where
\begin{align}
   &\lambda_1= -\frac{4\, \Gamma(1/\eta)}{\Gamma\big(1-1/\eta\big)}
             \bigg[ \frac{\Gamma\big(1+\frac{\eta}{2-2\eta}\big)}
                         {2\sqrt{\pi} \Gamma\big(1+\frac{1}{2-2\eta}\big)}
             \bigg]^{2/\eta-2},\label{lambda1}\\
   &\lambda_+=\frac{1}{2\pi}\tan\Big(\frac{\pi}{2-2\eta}\Big)\, ,\\
       &\lambda_-=\frac{\eta}{12\pi}
              \frac{\Gamma\big(\frac{3}{2-2\eta}\big)\,
                    \Gamma^3\big(\frac{\eta}{2-2\eta}\big)}
                   {\Gamma\big(\frac{3\eta}{2-2\eta}\big)\,
                    \Gamma^3\big(\frac{1}{2-2\eta}\big)}\ .
\end{align}
%

\subsection{Vacuum spin-spin correlation functions}

At this stage,  without the knowledge of the  constants
occurring in expansions 
\eqref{dvtz} and \eqref{dvt+-}, it is already possible
to predict\footnote{It has been done for the first time in 
Ref.\cite{LutP75}.}
the exact values of the  exponents for the  vacuum spin-spin
correlation functions. They follow  immediately 
from the scaling dimensions of the fields occurring
in \eqref{dvtz}, \eqref{dvt+-} and \eqref{Heff}.

It may be worth recalling at this point,  that the  vacuum  sector
of the infinite XXZ chain is infinitely degenerate.
In general, the boundary conditions imposed on the finite critical 
chain do not preserve all the  global  symmetries discussed above, 
and they may
be spontaneously broken at the thermodynamic limit.
Here we consider only 
translational invariant vacuums with unbroken parity:
\begin{equation}\label{oiy}
     {\mathbb P}\, \ket{\vac}= {\mathbb K}\, \ket{\vac}= \ket{\vac}\ .
\end{equation}
To fulfill these requirements, we shall treat the infinite XXZ chain  as
the thermodynamic 
limit of finite chains  subject to  periodic boundary 
conditions. 
Then, different vacuum 
states can be distinguished by means of the
operator $S_z$\ \eqref{ups}:
\begin{equation}\label{oaldjku}
   S_{z}\, \ket{s}=s\, \ket{s}\, ,  \qquad  2 s\in {\mathbb Z}\ . 
\end{equation}
Also, by a  proper   choice of   phase factors, 
one can always set up the conditions
\begin{equation}\label{sder}
     {\mathbb C}\, \ket{s}= | -s\, \rangle\  \qquad\ \text{and}\ \qquad\ \ 
     {\mathbb T}\, \ket{s}= | -s\, \rangle\ .
\end{equation}
In the continuous limit, the vacuum\ $\ket{s}$  
flows toward the conformal primary state with  right
and  left conformal dimensions equal to $d_{s,0}/2$,
where $d_{s,n}$ is given by\ \eqref{dimen}\ 
\cite{WoyE87,Batchelor,Alcaraz}. 
This implies in particular that, for the 
XXZ chain with a finite number of sites $N\gg 1$,
the difference of  vacuum  energies corresponding to the
states $\ket{s}$ and  $\ket{0}$ is  $2\pi d_{s,0}/N+o(N^{-1})$.

{}From our previous analysis,
one can now predict  the following asymptotic 
expansions for the time-ordered  correlation functions:
\begin{align}
   &\bracket{ T\, \sigma^x_{l+j}(t)\, \sigma_j^x(0) } 
        \sim \frac{A}{ (l_+l_-)^{\frac{\eta}{2}}}\, 
      \bigg\{\, 1-\frac{B}{(l_+l_-)^{\frac{2}{\eta}-2}}
         +O\big(l^{-2}\log l, l^{8-8/\eta}\big)\,
           \bigg\}\nonumber\\ 
   &\qquad
          - \frac{ (-1)^l\ {\tilde A}}
                 { (l_+l_-)^{\frac{\eta}{2}+\frac{1}{2\eta}}}\,       
      \bigg\{\,\frac{1}{2}\, \Big(\,\frac{l_+}{l_-}+\frac{l_-}{l_+}\,\Big)
              +\frac{{\tilde B}}{(l_+l_-)^{\frac{1}{\eta}-1}}
              + O\big(l^{-2}\log l, l^{4-4/\eta}\big)\, \bigg\}+\ldots\ ,
      \label{corr-+}\\ 
   &\bracket{T\, \sigma^z_{l+j}(t)\, \sigma_j^z(0)   }
        \sim -\frac{1}{\pi^2\eta\ l_+l_-}\, 
        \bigg\{\, \frac{1}{2}\, 
\Big(\, \frac{l_+}{l_-}+\frac{l_-}{l_+}\,\Big)
                  \nonumber\\  
   &\qquad
         +\frac{{\tilde B}_z}{(l_+l_-)^{\frac{2}{\eta}-2}}\, 
       \bigg(\, 1+\frac{2-\eta}{4(1-\eta)}\, 
              \Big(\, \frac{l_+}{l_-}+\frac{l_-}{l_+}\,\Big)\, \bigg)
         + O\big(l^{-2}\log l, l^{8-8/\eta}\big)\, \bigg\}
                  \nonumber\\ 
   &\qquad
         +\frac{(-1)^l\ A_z}{ (l_+l_-)^{\frac{1}{2\eta}}}\,
         \bigg\{\, 1-\frac{B_z}{(l_+l_-)^{\frac{1}{\eta}-1}}
         + O\big(l^{-2}\log l, l^{4-4/\eta}\big)\, \bigg\}+\ldots\ ,
   \label{corrzz}
\end{align}
where
\begin{equation}\label{msdshg}
   l_{\pm}=l\pm\frac{t}{\varepsilon}\gg 1\ .
\end{equation}
The coefficients in the asymptotic expansions\ 
\eqref{corr-+} and \eqref{corrzz}
do not really  depend on the choice of the vacuum
state $\ket{s}$, thus any finite $s\in {\mathbb Z}/2 $ can be chosen
for  the averaging
$\bracket{\ldots}\equiv\frac{\bracket{ s\, |\, \ldots\, |\, s}}{
\bracket{ s\, |\, s}}$.

In Eqs.~\eqref{corr-+}, \eqref{corrzz}, the correlation amplitudes 
$A$, ${\tilde A}$ and $A_z$ are simply related to
the first constants
occurring in the expansions\ \eqref{dvtz} and  \eqref{dvt+-}: 
\begin{equation}\label{amplc}
   A=2\,(C_0)^2\, ,\qquad
   {\tilde A}=(C_1)^2\, ,\qquad
   A_z=\frac{1}{2}\ (C_1^z)^2\ .
\end{equation}
They  will be computed in the next section.
At the same time,
the constants $B,\, {\tilde B},\, B_z$\ and\ $ {\tilde B}_z$ appearing in
\eqref{corr-+} and \eqref{corrzz}\ 
can be determined by
methods of conformal perturbation theory\ \cite{ZamAl}
based on the effective Hamiltonian\ \eqref{Heff}. 
In particular, the constant $B$
was  obtained in Ref.~\cite{Luk98} from
second order perturbative calculations,
\begin{equation}
  B=\frac{\lambda_1^2}{ 16}\ 
   \bigg\{\, \frac{2\pi^2}{\sin^2(2\pi/\eta)}
         -\frac{\eta^2}{(1-\eta)(2-\eta)}-\psi'(1/\eta)
         -\psi'(3/2-1/\eta)\, \bigg\},
\end{equation}
where $\psi'(z)=\partial_z^2\log\Gamma(z)$ and $\lambda_1$ is given by
\eqref{lambda1}.
The constant ${\tilde B}_z$ can be computed
similarly:
\begin{equation}
     {\tilde B}_z=\frac{\lambda_1^2}{4}\ 
\frac{\eta}{(\eta-2)^2}\  .
\end{equation}
On the contrary, the determination
of  ${\tilde B}$ and  $B_z$ do not  require
calculations beyond the   first order. They read explicitly,
\begin{align}
     &B_z=-\lambda_1\ 2^{\frac{4}{\eta}-5}\ 
          \frac{\Gamma(\frac{1}{\eta}-\frac{1}{2})\,
                \Gamma(1-\frac{1}{\eta})}
               {\Gamma(\frac{3}{2}-\frac{1}{\eta})\, 
                \Gamma(\frac{1}{\eta})}\ ,\\
     &{\tilde B}=(1-\eta)^2\ B_z\ .
\end{align}

\section{Calculation of correlation amplitudes}
\label{secDVT}

The purpose of this section is to explain how the local  
operators $\sg_l^\pm$ and $\sg_l^z$ can be quantitatively related 
to the scaling fields \eqref{dPhi}
and \eqref{op}, that is, how one can compute 
explicit analytic expressions for the constants $C_m$ and $C_m^z$ 
occurring in \eqref{dvtz} and \eqref{dvt+-} for the fixed CFT normalization 
\eqref{CFTnorm}. 
We concentrate here on the 
first terms of the expansions \eqref{dvtz}, \eqref{dvt+-}, which provide the
main  asymptotic behavior of the spin-spin correlation functions
\eqref{corr-+} and  \eqref{corrzz}:
\begin{align}
   &\sg^\pm_l \sim  C_0\, \eps^{\frac{\eta}{2}}\ \cO_{\pm 1, 0} (x) 
               + (-1)^l\  \frac{C_1}{2} \, 
                 \eps^{\frac{\eta}{2}+\frac{1}{2\eta}}\ 
                 \big(\cO_{\pm 1,2} +\cO_{\pm 1, -2}\big)(x)+\dots\ ,
                       \label{dvt-spin2}\\
   &\sg^z_l \sim \frac{\eps}{2\pi\sqrt{\eta}}\  \partial_t\varphi(x)
             +  (-1)^l\  \frac{C^z_1}{2i }\, \eps^{\frac{1}{2\eta}} \, 
                \big( \cO_{0,2} -\cO_{0, -2} \big)(x)
             +\dots\ .\label{dvt-spin1}
\end{align}
Note that in \eqref{dvt-spin1} each of the
two terms is either leading or sub-leading according to the value of  
$\eta$, i.e. of the anisotropy parameter $\Delta$, whereas in 
\eqref{dvt-spin2} the second term is always subleading. Nevertheless, this
term  gives rise to noticeable numerical 
corrections to the leading asymptotic
behavior of correlation functions and has already been studied
numerically (see Section~\ref{secASY}).

\subsection{Even sector of the infinite XYZ chain}

To find the  quantitative relation between  the spin operators 
and the local scaling  fields,
it is useful to move slightly away from
criticality and, instead of \eqref{Ham}, to consider the XYZ chain,
\begin{equation}\label{HXYZ}
   \H_{\mathrm{XYZ}}=-\frac{1}{2} \sum_{l=-\infty}^{\infty}
           \Bigl\{ J_x\,  \sg_l^x \sg_{l+1}^x 
        + J_y\, \sg_l^y \sg_{l+1}^y + J_z\, \sg_l^z \sg_{l+1}^z 
 \Bigr\}\ . 
\end{equation}
Without loss of generality we assume here  that
$J_x > J_y\geq |J_z|.$
The XYZ deformation has  a remarkable feature:
it preserves the integrability of the original  theory
\cite{Bax82L,Baxter2}.
Nowadays, the structure of the
Hilbert space of the infinite  XYZ chain is  well understood.
We recall here some basic facts that will be useful for our analysis.

In the case of the XYZ spin chain, the global
symmetry group discussed in the previous section is  
explicitly broken to the subgroup generated 
by the lattice translation ${\mathbb K}$,
the  ${\mathbb C}$, ${\mathbb P}$, ${\mathbb T}$ transformations
and the rotation\ \eqref{Ulatt}\ with 
$\alpha=\pi$: ${\mathbb U}_{\pi}$.
In this section, we  concentrate on the even sector of the
XYZ spin chain 
(defined as the  thermodynamic limit of  finite chains with 
an  even number of sites),
which  implies the condition
\begin{equation}\label{quas}
                  {\mathbb U}^2_{\pi}=1\ .
\end{equation}

Let  us  denote by ${\cal H}_s$ the  eigenspace of the operator
$S_z$ corresponding to  a given eigenvalue $s\in {\mathbb Z} $. 
Then, the   Hamiltonian\ \eqref{HXYZ}\
acts as,
\begin{equation*}
    \H_{\mathrm{XYZ}}\, :\, {\cal H}_s\to {\cal H}_{s-2}\oplus
                            {\cal H}_{s}\oplus {\cal H}_{s+2}\ ,
\end{equation*}
and therefore the infinite degeneracy
in the vacuum sector of the XXZ chain is reduced to the two states 
\begin{equation*}
    \ket{s}_{\mathrm{XYZ}}\in 
    \oplus_{k=-\infty}^{\infty}\, {\cal H}_{s+2 k} \qquad (\, s=0, 1\, )
\end{equation*}
satisfying the condition\footnote{
The corresponding energies  $E^{(s)}_N$ for a chain
with a finite number of sites $N\gg 1$ are
asymptotically degenerate in the sense that 
$E^{(1)}_N-E^{(0)}_N=O\big(e^{-\text{const}\,  N}\big)$\ 
(see e.g. Ref.\cite{Bax82L}).}
\begin{equation}\label{possy}
    {\mathbb U_{\pi}}\, \ket{s}_{\mathrm{XYZ}}=
          e^{i\pi s}\  \ket{s}_{\mathrm{XYZ}}\ .
\end{equation}
This requirement, along with the conventional normalization
of  vacuum states, 
$\langle\, {\vac }\, |\, {\vac }\, \rangle=1$, defines 
$\ket{s}_{\mathrm{XYZ}}$ up to an  overall complex  phase factor.
Since the  time reversal
transformation acts on  states  as the complex conjugation,
one can eliminate (up to sign)  the  ambiguity of such a phase
by imposing the condition
\begin{equation*}
    {\mathbb T}\, \ket{s}_{\mathrm{XYZ}}= \ket{s}_{\mathrm{XYZ}}\ .
\end{equation*}

To gain physical intuition about the vacuum sector,
it is useful to consider a limiting case 
where the Hamiltonian\ \eqref{HXYZ}\ simplifies drastically.
For $J_y=J_z=0$, the vacuum sector of the finite periodic XYZ chain
contains two pure ferromagnetic states 
$\ket{\vac}^{(j)}$ ($j=0,1$).
With a proper choice of the overall  phases 
of these states,  one may always set up the
conditions
\begin{equation}\label{refq1}
  {\mathbb U}_{\pi}\, \ket{\vac}^{(j)}=\ket{\vac}^{(1-j)}\, , \qquad\quad  
  {\mathbb T}\,  \ket{\vac}^{(j)}=\ket{\vac}^{(j)}\ .
\end{equation}
Moreover, since the charge conjugation matrix\ \eqref{trPlat}\ 
can be identify with $\prod_{l}\sigma_l^x$, one has
\begin{equation}\label{refq34}
  {\mathbb C}\, \ket{\vac}^{(j)}=\ket{\vac}^{(j)}\ .
\end{equation}
When the couplings $J_y$ and $J_z$ are non-vanishing, 
the  pure ferromagnetic states are no longer stationary
states, but it is still possible to introduce two vacuums
$\ket{\vac}^{(j)}$ satisfying
\eqref{refq1} and \eqref{refq34}
through the relation
\begin{equation}
   \ket{s}_{\mathrm{XYZ}}=\frac{1}{\sqrt{2}}\ 
   \big\{\, \ket{\vac}^{(0)} +(-1)^s\, \ket{\vac}^{(1)}\, \big\}
   \qquad\,  (s=0, 1)\ .
\end{equation}

The   Hilbert space of the XYZ chain contains two linear
subspaces ${\cal V}^{(j)}$ ($j=0,1$) associated with the vacuums  
$\ket{\vac}^{(j)}$.
In the spectrum of the model, 
there exist   kink-like ``massive'' excitations,
${\bf B}_{+}$ and ${\bf B}_{-}$, such that the corresponding
Zamolodchikov-Faddeev operators  intertwine these 
subspaces  ${\cal V}^{(j)}$:
\begin{equation*}
    {\bf B}_{\pm}\, :\ {\cal V}^{(j)}\to {\cal V}^{(1-j)}\ .
\end{equation*}
One  can therefore generate two sets of asymptotic states in the form
\begin{equation}\label{states}
       {\bf B}_{\sigma_{2m}}(k_{2m})\ldots {\bf B}_{\sigma_{1}}(k_{1})\,
       \ket{\vac}^{(j)}\in {\cal V}^{(j)}\,  \qquad\  (m=0,1, 2\ldots)\ ,
\end{equation}
%
where the  Zamolodchikov-Faddeev  operators
${\bf B}_{\sigma}(k)\ (\sigma=\pm)$  depend on
a quasi-momentum $k$:
\begin{equation}\label{transaa}
    {\mathbb K}\, {\bf B}_{\pm}(k)\, {\mathbb K}^{-1}=
         e^{ik}\ {\bf B}_{\pm}(k)\, ,
           \qquad\quad 
    \big[\, \H_{\mathrm{XYZ}}\, ,\, {\bf B}_{\pm}(k)\, \big]=
         {\cal E}(k)\ {\bf B}_{\pm}(k)\ .
\end{equation}
The dispersion relation ${\cal E}={\cal E}(k)$ of the fundamental
excitations was calculated in work\ \cite{JohKM73}. It is recalled 
in Appendix A (see Eqs.~\eqref{poi}, \eqref{poii}). 
The operators ${\bf B}_{\pm}$
satisfy also the  conditions:
\begin{equation}\label{trans}
   {\mathbb C}\, {\bf B}_{\pm}^{(1-j,j)}\, {\mathbb C}=
      \mp\, (-1)^{j}\ {\bf B}^{(1-j,j)}_{\pm}\, ,
              \qquad\quad 
   {\mathbb U}_{\pi}\, {\bf B}^{(1-j,j)}_{\pm}\, {\mathbb U}_{\pi}=
          \pm\, (-1)^{j}\   {\bf B}^{(j,1-j)}_{\mp}\  .
\end{equation}
In Eqs.~\eqref{trans}, ${\bf B}^{(1-j,j)}_{\pm}$ denotes 
the restriction of the  operator ${\bf B}_{\pm}$ when it acts
on the subspace ${\cal V}^{(j)}$.

Any local  lattice  operator ${\cal O}_{\rm latt}$
(i.e. any  operator which can be written as
a local combination of $\sigma_l^a$)
leaves the subspaces ${\cal V}^{(j)}$ invariant:
\begin{equation*}
   {\cal O}_{\rm latt}\, :\, {\cal V}^{(j)}\to {\cal V}^{(j)}\ .
\end{equation*}
Furthermore, 
the   algebra of local operators ${\cal A}_{\rm loc}$ acts 
invariantly on the 
component  of the  subspace ${\cal V}^{(j)}$ generated
by the  states\ \eqref{states}.  
For $J_z\leq 0$, the kinks  ${\bf B}_{\pm}$ do not produce bound
states, and the two sets of 
asymptotic states\ \eqref{states}\ obeying the condition
\begin{equation*}
        -\pi/ 2\leq  k_1<k_2\ldots <k_{2m} < \pi/ 2
\end{equation*}
form complete in-bases in  these  unitary equivalent 
spaces of representation of ${\cal A}_{\rm loc}$.
At first glance, to construct explicitly
the representations of ${\cal A}_{\rm loc}$, 
one should  put at one's disposal the whole  collection of
in-basis matrix elements for  an arbitrary local operator
${\cal O}_{\text{latt}}\in {\cal A}_{\rm loc}$.
As a matter of fact, using the
so-called crossing symmetry\ (see e.g.\cite{JimM95L,Smirnov}),
one can express all
possible   matrix elements in terms of those of the form
\begin{equation}\label{jshd}
  {}^{(j)}\bra{\vac}\, {\cal O}_{\text{latt}}\,  
     {\bf B}_{\sigma_{2m}}(k_{2m})\ldots {\bf B}_{\sigma_{1}}(k_{1})\,
           \ket{\vac}^{(j)} 
    \equiv
  {}^{(j)}\big\langle\, {\cal O}_{\text{latt}}\, |\, 
  {\bf B}_{\sigma_{1}}(k_{1})\ldots {\bf B}_{\sigma_{2m}}(k_{2m})\,
  \big\rangle_{\rm in}\, .
\end{equation}
Such matrix elements are known as  form factors.

Currently, there  exists a formal 
procedure which allows one
to express   form factors of local operators in 
terms of multiple integrals\ \cite{LasP98,Las01}.
Unfortunately,
it is difficult to apply in practice,  
even in the case of the local spin operators
$\sigma^a_l$ themselves  and for
form factors involving only a small number of excitations.
For our  purposes, we merely need the explicit form of
Vacuum Expectation Values (VEV) and of two-particle form factors.
{}From the relations\ \eqref{oiy}, \eqref{refq1} and \eqref{refq34},
it follows immediately 
that the VEVs of $\sigma^y_l$ and $\sigma^z_l$ 
vanish, whereas
\begin{equation}\label{vxcd}
    {}^{(j)}\big\langle\,  \sigma^x_l\,\big\rangle=(-1)^j\ F\ ,
\end{equation}
where $F$ depends on the two ratios 
$J_y/J_x$ and $J_z/J_x$.
This VEV  was found in work\ \cite{BaxK74}
(see Appendix A, Eq.~\eqref{B-K}).
As for the two-particle form factors,
the ${\mathbb Z}_2$-symmetry generated
by  ${\mathbb U}_{\pi}$\ \eqref{trans}\ enables one to predict 
their general form: 
\begin{align}\label{wtsr}
   &{}^{(j)}\big\langle\, \sigma^x_0\, |\,
         {\bf B}_{\pm}(k_{1}) {\bf B}_{\mp}(k_{2})\big\rangle_{\rm in}=
         (-1)^j\  F^x_1(k_1,k_2)\pm F^x_2(k_1,k_2)\, ,\\
            \label{jsshyb}
   &{}^{(j)}\big\langle\,  \sigma^y_0\, |\, 
         {\bf B}_{\pm}(k_{1}) {\bf B}_{\pm}(k_{2})\big\rangle_{\rm in}=
          F^y_1(k_1,k_2)\pm (-1)^j\ F^y_2(k_1,k_2)\, ,\\ 
            \label{jsshyb1}
   &{}^{(j)}\big\langle\,  \sigma^z_0\,|\, 
         {\bf B}_{\pm}(k_{1}) {\bf B}_{\pm}(k_{2})\big\rangle_{\rm in}=
         (-1)^j\  F^z_1(k_1,k_2)\pm  F^z_2(k_1,k_2)\, ,
\end{align}
where $ F^a_{1,2}$ are some functions of the quasi-momentums $k_1$ and
$k_2$.
The invariance with respect to charge conjugation 
${\mathbb C}$\ \eqref{trans}\ dictates that all  other
two-particle form factors vanish.
Form factor\ \eqref{wtsr}\ was calculated in 
Ref.\cite{Las01}. 
It  is possible to extend the result of that work and calculate
form factors\ \eqref{jsshyb} and \eqref{jsshyb1} as well.
We collect the explicit expressions of  $F$ and $F^a_{1,2}$
in Appendix A.

To conclude this subsection,
let us note that, in the current  treatment
of  the infinite  XYZ model as the thermodynamic
limit of  finite  periodical chains  with an even
number of sites,  we cannot construct   excited states
containing  an odd number of the elementary excitations
\begin{equation}\label{statesa}
      {\bf B}_{\sigma_{2m+1}}(k_{2m+1})
\ldots {\bf B}_{\sigma_{1}}(k_{1})\,
      |\, {\rm vac}\rangle^{(1-j)}\in {\cal V}^{(j)}\,  \ \ \ \ \quad\ \
      (m=0,1\ldots)\ .
\end{equation} 
Such states are deduced  from finite chains
with boundary conditions breaking the translation
invariance. In particular, the linear subspace of  ${\cal V}^{(j)}$
spanned by the states\ \eqref{statesa}\ can be constructed from the
thermodynamic limit
of   chains with an odd number
of sites  $N$ and
subject to the so called twist boundary condition\footnote{This 
fact follows from the result of work\ \cite{Alcaraz}. }:
\begin{equation*}
     \sigma_{-\frac{N-1}{2 }}^{\pm}=-\sigma_{\frac{N+1}{2 }}^{\pm}\, ,
           \qquad\ \
     \sigma_{-\frac{N-1}{2 }}^{z}=\sigma_{\frac{N+1}{2 }}^{z}\   .
\end{equation*}
The  in-asymptotic states\ \eqref{states}, \eqref{statesa}
form
complete bases in  ${\cal V}^{(j)}$ for $J_z\leq 0$.

\subsection{XYZ spin chain in the scaling limit}

For $J_x= J_y$, the gap in the spectrum of the XYZ chain
vanishes, and its correlation length~\cite{Lut76}
\begin{equation}\label{nxbt}
   R_c\simeq \frac{1}{4}\  \bigg[
       \frac{8\, (  J^2_x-J_z^2)}{J_x(J_x-J_y)}\bigg]^{\frac{1}{2-2\eta}}\, 
   \quad\qquad 
   \Big(\eta=\frac{1}{\pi}\, \arccos(J_z/J_x)\, \Big) 
\end{equation}
becomes infinite. In the
limit $J_x\to J_y$, the correlation functions at large lattice
separation ($\sim R_c$)
assume a certain scaling form
which can be described by quantum field theory.
If $(J_x-J_y)/J_x\ll 1$, it is natural to  treat the XYZ
model  as the  perturbation
of the XXZ chain by the lattice operator
\begin{equation*}
    \sigma^x_l\sigma^x_{l+1}-\sigma^y_l\sigma^y_{l+1}\ .
\end{equation*}
The leading term in 
the expansion \eqref{corrDVT} for  this  operator
is given by the  relevant  field ${\cal O}_{2,0}+{\cal O}_{-2,0}$
of  scaling dimension $2\eta$. 
Therefore, the scaling limit of the XYZ chain is described by 
the  sine-Gordon quantum field theory\ \cite{Lut76},
\begin{equation}\label{SineG}
   {\bf H}_{\mathrm{sg}}={\bf H}_{\mathrm{Gauss}}-\mu\, \int dx\, 
           \big({\cal O}_{2,0}+{\cal O}_{-2,0}\big)(x)\ .
\end{equation}
Up to a  numerical factor, which was obtained in Ref.~\cite{Zam95},
the coupling constant $\mu$ coincides with the quantity $M^{2-2\eta}$,
where the combination
\begin{equation}\label{mass}
     M=( \varepsilon R_c)^{-1}
\end{equation}
can be naturally identified with the soliton mass in the sine-Gordon 
model\ \eqref{SineG}.

The sine-Gordon theory
admits a class of local soliton-creating operators  characterized
by two integers $s,n\in\Zset$, where
$n$ gives the topological charge and $sn/2$ represents
the  Lorentz spin  of the field.
These operators can also be expressed in a form similar to  \eqref{op},
in which $\varphi$ denotes the sine-Gordon field 
instead of the  Gaussian field 
obeying the simple D'Alembert equation\ \eqref{vsfdr} 
(see e.g. Ref.~\cite{LukZ01} for details).
They moreover coincide with the Gaussian fields\ \eqref{op} 
in the conformal  limit $\mu\tend 0$. Hence, with some abuse
of notation, we will  denote 
such soliton-creating operators in the sine-Gordon model by the
same symbol  $\cO_{s, n}$\footnote{In Ref.~\cite{LukZ01}, 
the  soliton-creating operators
were denoted as $\cO_a^n$,  where
$a=s\sqrt{\eta}/2$. The parameter $\beta$ in \cite{LukZ01} 
coincides with $\sqrt{\eta}$. 
Note that there, the quantity $2 a/\beta$ was
not assumed to be integer.}.

To proceed further, one needs to draw a link
between local lattice operators in the XYZ chain
and local fields in the sine-Gordon model. 
Let us note at this point that local   expansions of the type\
\eqref{corrDVT} 
are based on dimensional analysis and do not necessarily imply
the criticality of the original  lattice system.
Similar expansions are   expected to be applicable
to describe the  near-critical behavior  of  lattice systems.
Usually, the  term with 
the smallest scaling dimension in \eqref{corrDVT}\ 
governs the  universal scaling behavior of
lattice  correlators, whereas
the next terms   produce non-universal corrections.
In particular, 
relations\ \eqref{dvt-spin2}, \eqref{dvt-spin1}
obtained for the  XXZ spin  chain
can be used to study  the  leading  scaling  behavior and
first  non-universal 
corrections of the  XYZ  correlation functions and form factors.
In the XYZ case, the   continuous fields which appear in
\eqref{dvt-spin2}, \eqref{dvt-spin1}
should be understood 
as operators in  the sine-Gordon model rather 
than  their conformal limits.
The numerical constants $C_0$, $C_1$ and  $C_1^z$
remain, of course, the same as for the critical XXZ chain.

Let us now discuss the relation between the  Hilbert spaces of
the XYZ and sine-Gordon models. In general, the theory\ \eqref{SineG}\
admits a discrete symmetry $\varphi\to \varphi+2\pi j/\sqrt{\eta}$
$(j\in  {\mathbb Z})$, which is generated
by the operator ${\mathbb U}_{\pi}$ defined similarly to
\eqref{SZZA}.
For $0<\eta\leq 1$, the above symmetry 
is spontaneously broken, so that 
the theory has an infinite number of ground 
states\ $\ket{ 0_j}$  $(j\in  {\mathbb Z}) $
characterized by the corresponding VEVs of the
field $\varphi$:
\begin{equation}\label{kjhgnm}
   \frac{\bra{ 0_j}\, \varphi(x)\,\ket{0_j}}
        {\langle \, 0_j\, |\, 0_j\, \rangle}
   =\frac{2\pi j}{\sqrt{\eta}}\ .
\end{equation}
The sine-Gordon model which governs the scaling behavior of the even
sector of the XYZ chain is subject to the additional constraint 
${\mathbb U}_{\pi}^2=1$.
This equation implies in particular 
that the field $\varphi$ is compactified, 
$\varphi\equiv\varphi+4\pi/\sqrt{\eta}$, and that,
unlike the uncompactified case, 
there exist only two non-equivalent vacuum states
$|\, 0_j\, \rangle$ with
$j=0,1$. These states  are naturally identified with the scaling
limit of the two XYZ  vacuums $|\, {\rm vac}\, \rangle^{(j)}$.

To describe the scaling limit of XYZ excited states, 
one has to relate
the Za\-mo\-lodchi\-kov-Faddeev operators of the lattice and continuous
theories. Let us recall here that the sine-Gordon model
admits a global continuous $U(1)$ symmetry generated by 
the operator
\begin{equation}
\label{SZaZA}
   {\mathbb V}_{\alpha}=e^{i\alpha Q}\, , \quad\ {\rm where}\  \quad
      Q=\frac{ \sqrt{\eta}}{2\pi}\,
                \int_{-\infty}^{\infty}dx\, \partial_{x}\varphi\ ,
\end{equation}
which  acts on the exponential fields as follows,
\begin{equation*}
   {\mathbb V}_{\alpha}\, {\cal O}_{s,n}\, {\mathbb V}^{-1}_{\alpha}=
   e^{i n\alpha}\ {\cal O}_{s,n}\ .
\end{equation*}
Notice that the Gaussian CFT
also possesses  such a global symmetry,
contrary to the  XXZ and XYZ lattice models.
Nevertheless, the form of the expansions \eqref{dvtz}, \eqref{dvt+-}
and\ \eqref{nhst}\ suggests that
the lattice models are   invariant  with respect to the
transformation $ {\mathbb  V}_{\pi}$ which 
acts trivially on
all local lattice  fields.
Such symmetry manifests 
itself in the existence of two subspaces  ${\cal V}^{(j)}$
($j=0,1$) which can be treated as  eigenspaces 
of the operator  $ {\mathbb V}_{\pi}$:
\begin{equation*}
   {\mathbb  V}_{\pi}\,  {\cal V}^{(j)}=(-1)^j\  {\cal V}^{(j)}\ .
\end{equation*}
The fundamental  sine-Gordon kink-like excitations, 
the soliton  ${\bf A}_-$ and the antisoliton ${\bf A}_+$, 
carry,  respectively, negative and positive units 
of the  topological charge $Q$\ \eqref{SZaZA}:
\begin{equation}\label{pais}
   {\mathbb V}_{\alpha}\, 
{\bf A}_{\pm}(\theta)\, {\mathbb V}^{-1}_{\alpha}=
   e^{\pm i\alpha}\ {\bf A}_{\pm}(\theta)\ ,
\end{equation}
where the argument $\theta$ denotes kink rapidity.
The relation with the Zamolodchikov-Faddeev operators of the XYZ 
model was established in Ref.\cite{Miwa}. In our notations
it can be summarized as follows: 
the operators ${\bf A}_{\pm}(k)$ defined 
in the lattice  model as
\begin{equation}\label{jshyt}
   {\bf A}^{(1,0)}_{\pm}=\frac{i}{\sqrt{2}}\, 
        \big(\, \pm{\bf B}^{(1,0)}_{+}+ {\bf B}^{(1,0)}_{-}\,\big)\, ,
   \qquad\ 
   {\bf A}^{(0,1)}_{\pm}=\frac{i}{\sqrt{2}}\, 
       \big( -  {\bf B}^{(0,1)}_{+} \pm {\bf B}^{(0,1)}_{-}\,\big)\, ,
\end{equation}
turn out, in the scaling limit, to be the Zamolodchikov-Faddeev 
operators ${\bf A}_{\pm}(\theta)$ \eqref{pais} of the sine-Gordon model.
Here, as well as  in Eq.~\eqref{trans}, we denote the restriction of the  
operators ${\bf B}_{\pm}$ and ${\bf A}_{\pm}$ acting
on the subspace ${\cal V}^{(j)}$ as ${\bf B}^{(1-j,j)}_{\pm}$ and
${\bf A}^{(1-j,j)}_{\pm}$.
Again, with some abuse of notation, we use the same symbol
${\bf A}_{\pm}$ for the XYZ operators and for their scaling limits.
Notice that  the quasi-momentum $k$ of
the low-lying fundamental  excitation  becomes the
usual particle  momentum in the scaling  limit:
\begin{equation*}
   \lim_{\varepsilon\to 0}\, \frac{k}{\varepsilon}= M\, \sinh(\theta)
   \quad \text{and}\quad 
   \lim_{\varepsilon\to 0}\, \frac{{\cal E}(k)}{\varepsilon}= 
               M\, \cosh(\theta)\, ,
\end{equation*}
where  $M$ is  the soliton mass \eqref{mass}.

\subsection{Scaling behavior of the form factors}

We are now in a position to calculate  the
constants appearing in  
expansions\ \eqref{dvt-spin2}, \eqref{dvt-spin1}.
To illustrate the procedure, 
let us consider first the scaling behavior of the lattice 
VEV\ \eqref{vxcd}. From relation
\eqref{dvt-spin2}\ we deduce that
\begin{equation}\label{zdsyhnyh}
   F\sim\, 2\, C_0\, \varepsilon^\frac{\eta}{2}\, 
           \big\langle \, {\cal O}_{1,0}\, \big\rangle+\ldots\ .
\end{equation}
Here $\langle\, \ldots\,   \rangle$ means an averaging with respect to the
vacuum\ \eqref{kjhgnm} with $j=0$. To write the equation\ \eqref{zdsyhnyh}, 
we also use the fact that the VEV
of the fields $ {\cal O}_{1,0}$ and  $ {\cal O}_{-1,0}$ are equal 
by virtue of charge conjugation symmetry.
The VEV of the operator 
${\cal O}_{s,0}$ was found in\ \cite{LukZ97}. We denote it
as
\begin{equation*}
   \big\langle \, {\cal O}_{s,0}\, \big\rangle=\sqrt{{ Z}_{s,0}}\ .
\end{equation*}
On the other hand, 
the function $F$ is given by the Baxter-Kelland formula
\cite{BaxK74} (see Eq.\eqref{B-K}).
Its leading scaling behavior  reads
\begin{equation}\label{vac-ss}
   F\sim \frac{1}{1-\eta}\, (4R_c)^{-{\frac{\eta}{2}}}+\ldots\, .
\end{equation}
Comparing\ Eqs.~\eqref{zdsyhnyh}\ and\ \eqref{vac-ss}, 
and using the relation
$R^{-1}_c=M\varepsilon$, one can deduce  the constant $C_0$: 
\begin{equation}\label{22}
  C_0=\frac{1}{2(1-\eta)\sqrt{{{Z}}_{1,0}}}\,
                \Big(\frac{M}{4}\Big)^{\frac{\eta}{2}}\, .
\end{equation}

A similar strategy can be applied to calculate the constants $C_1$ and
$C_1^z$ in\ \eqref{dvt-spin2}, \eqref{dvt-spin1}.
The relations\ \eqref{jshyt}\ allows one to express
the leading scaling behavior of the
functions that appear in Eqs.~\eqref{wtsr}-\eqref{jsshyb1}\
through the form factors of the
sine-Gordon fields:
\begin{align}\label{sndg}
   & F_1^x\pm i\,   F_1^y \sim  2\, C_0\, \varepsilon^\frac{\eta}{2}\,
          \big\langle\, {\cal O}_{\pm 1,0}(0)\,
          |\, {\bf A}_{+}(\theta_1){\bf A}_{-}(\theta_2)\,
          \big\rangle_{\rm in}+\ldots\, ,\\
             \label{sndg1}
   & F_2^x\pm i\, F_2^y 
        \sim - C_1\, \varepsilon^{\frac{\eta}{2}+\frac{1}{2\eta}}\, 
        \big\langle\, {\cal O}_{\pm 1,2}(0)\, 
          |\, {\bf A}_{-}(\theta_1){\bf A}_{-}(\theta_2)\,
          \big\rangle_{\rm in}+\ldots\, ,
\end{align}
and
\begin{align}\label{sndg2}
   &F_1^z\sim  \frac{  \varepsilon}{2\pi\, \sqrt{\eta}}\ 
         \big\langle\, \partial_{t}\varphi(0)\, 
         |\, {\bf A}_{+}(\theta_1){\bf A}_{-}(\theta_2)\, 
         \big\rangle_{\rm in}+\ldots\, ,\\
             \label{sndg3}
   &F_2^z\sim  \frac{ i}{2 }\ C_1^z\  \varepsilon^{\frac{1}{2\eta}}\ 
         \big\langle\, {\cal O}_{0 ,2}(0)\, 
         |\, {\bf A}_{-}(\theta_1){\bf A}_{-}(\theta_2)\,
         \big\rangle_{\rm in}+\ldots\ .
\end{align}
We use here an abbreviated notation similar to\ \eqref{jshd},
except that the index specifying the vacuum states is omitted
since it is always assumed to be $j=0$.
The two-particle form factors of the topologically neutral operators
${\cal O}_{\pm 1,0}$ and $\partial_{t}\varphi$ have been known for
a long time (see e.g.\cite{Smirnov}). They read explicitly,
\begin{align}
   &\big\langle\, {\cal O}_{\pm 1,0}(0)\, 
      |\, {\bf A}_{+}(\theta_1){\bf A}_{-}(\theta_2)\,
      \big\rangle_{\rm in}
    =\sqrt{{ Z}_{1,0}}\ \frac{G(\theta_1-\theta_2)}{\xi\ G(-i\pi)}\ 
     \frac{2 i\ e^{\mp(\theta_1-\theta_2+i\pi)/(2 \xi) }}
          {\sinh\big((\theta_1-\theta_2+i\pi)/\xi\big)}\, ,\\
   &\big\langle\, \partial_{t}\varphi(0)\, 
      |\, {\bf A}_{+}(\theta_1){\bf A}_{-}(\theta_2)\, 
      \big\rangle_{\rm in}
    =\frac{G(\theta_1-\theta_2)}{\sqrt{\eta}\ G(-i\pi)}\ 
     \frac{i\pi \,  M\  \big(e^\frac{\theta_1+\theta_2}{2}+
           e^{-\frac{\theta_1+\theta_2}{2}}\big)}
          {\cosh\big((\theta_1-\theta_2+i\pi)/(2 \xi)\big)}\, ,
\end{align}
where $\xi=\frac{\eta}{1-\eta}$ and
the function $G(\theta)$ is a so-called minimal form factor.
The  form factors of
the topologically charged operators $\cO_{s,n}$ have been proposed in
\cite{LukZ01}. In particular, for $n=2$ one has:
\begin{equation}
   \big\langle\, {\cO}_{s, 2}(0)\, 
      |\,{\bf A}_{-}(\theta_1){\bf A}_{-}(\theta_2)\,
      \big\rangle_{\text{in}}
   =\sqrt{{ Z}_{s,2}}\ e^{\frac{i{\pi  s}}{{2}}}\,
        e^{\frac{s \theta_1}{2}+\frac{s\theta_2}{2}}\,
        G(\theta_1 - \theta_2)\, .
\end{equation}
The explicit  formulae for the minimal form factor $G(\theta)$ and
the field-strength renormalization ${ Z}_{s,n}$
are recalled in Appendix B.

We are now able to compare \eqref{sndg}-\eqref{sndg3}
with   expansions of the exact lattice form factors
\eqref{vxcd}-\eqref{jsshyb1}
given in Appendix A, and to relate
the values of the constants $C_0$, $C_1$ and
$C_1^z$ to the constants  ${ Z}_{s,n}$:
\begin{align}\label{2}
  &C_1=\frac{4}{\eta\,G(-i\pi)\, \sqrt{{{Z}}_{1,2}} }\,
       \Big(\frac{M}{4}\Big)^{\frac{\eta}{2}+\frac{1}{2\eta}}\, , \\
  \label{23} 
  &C_1^z=\frac{8}{\eta\, G(-i\pi)\, \sqrt{{{Z}}_{0,2} }}\,
       \Big(\frac{M}{4}\Big)^{\frac{1}{2\eta}}\, .
\end{align}
%

\section{Correlation amplitudes. Comparison with numerical results}
\label{secASY}

With the explicit expression \eqref{norm} for the normalization constants
${{Z}}_{1,0}$, ${{Z}}_{1,2}$ and ${{Z}}_{0,2}$, 
the relations\ \eqref{22}, \eqref{2} and \eqref{23}\  
lead to the following formulae 
for the correlation  amplitudes   $A$,
${\tilde A}$ and $A_z$\ \eqref{amplc}:
\begin{equation}\label{Ax}
   {A}=\frac{1}{2(1-\eta)^2}\, 
      \bigg[\frac{\Gamma(\frac{\eta}{2-2\eta})}
                 {2\sqrt{\pi} \Gamma(\frac{1}{2-2\eta})}
      \bigg]^{\eta}
   \exp\bigg\{-\int_0^{\infty}\frac{dt}{t} 
      \Big(\frac{\sinh(\eta t)}{\sinh(t)\cosh((1-\eta)t)}
           -{\eta}\, e^{-2 t}\, \Big) \bigg\}\, ,
\end{equation}
\begin{align}\nonumber
   &\espneg{\tilde A}=\frac{2}{\eta (1-\eta)}\, 
      \bigg[\frac{\Gamma(\frac{\eta}{2-2\eta})}
                 {2\sqrt{\pi} \Gamma(\frac{1}{2-2\eta})}
      \bigg]^{\eta+\frac{1}{\eta}}\\ 
    \label{tAx}
   &\ \times \exp\bigg\{-\int_0^{\infty}\frac{dt}{t} 
      \Big(\frac{\cosh(2 \eta t) e^{-2 t}-1}
                {2\sinh(\eta t)\sinh(t)\cosh((1-\eta)t)}
   +\frac{1}{\sinh(\eta t)}-
\frac{\eta^2+1}{\eta}\, e^{-2 t}\, \Big) \bigg\}\, ,
\end{align}
and
\begin{equation}\label{Az}
   A_z=\frac{8}{\pi^2}\, 
     \bigg[\frac{\Gamma(\frac{\eta}{2-2\eta})}
                {2\sqrt{\pi} \Gamma(\frac{1}{2-2\eta})}
     \bigg]^{\frac{1}{\eta}}
     \exp\bigg\{\int_0^{\infty}\frac{dt}{t}
       \Big(\frac{\sinh((2 \eta-1) t)}{\sinh(\eta t)\cosh((1-\eta)t)}
       -\frac{2\eta-1}{\eta}\, e^{-2 t}\, \Big) \bigg\}\, .
\end{equation}
Note that these amplitudes obey the simple relation:
\begin{equation}
   \frac{{\tilde A}}{A A_z}
      =\frac{\pi}{4}\, 
       \frac{ \Gamma^2\big(1+\frac{\eta}{2-2\eta}\big)}
            {\Gamma^2\big(\frac{3}{2}+\frac{\eta}{2-2\eta}\big)}\, .
\end{equation}
The correlation amplitude $A$ was already obtained in \cite{LukZ97}.
Our computations also  confirm the conjecture 
from \cite{Luk99} concerning the amplitude $A_z$.

\begin{figure}[h]
\centering
\includegraphics[width=12cm]{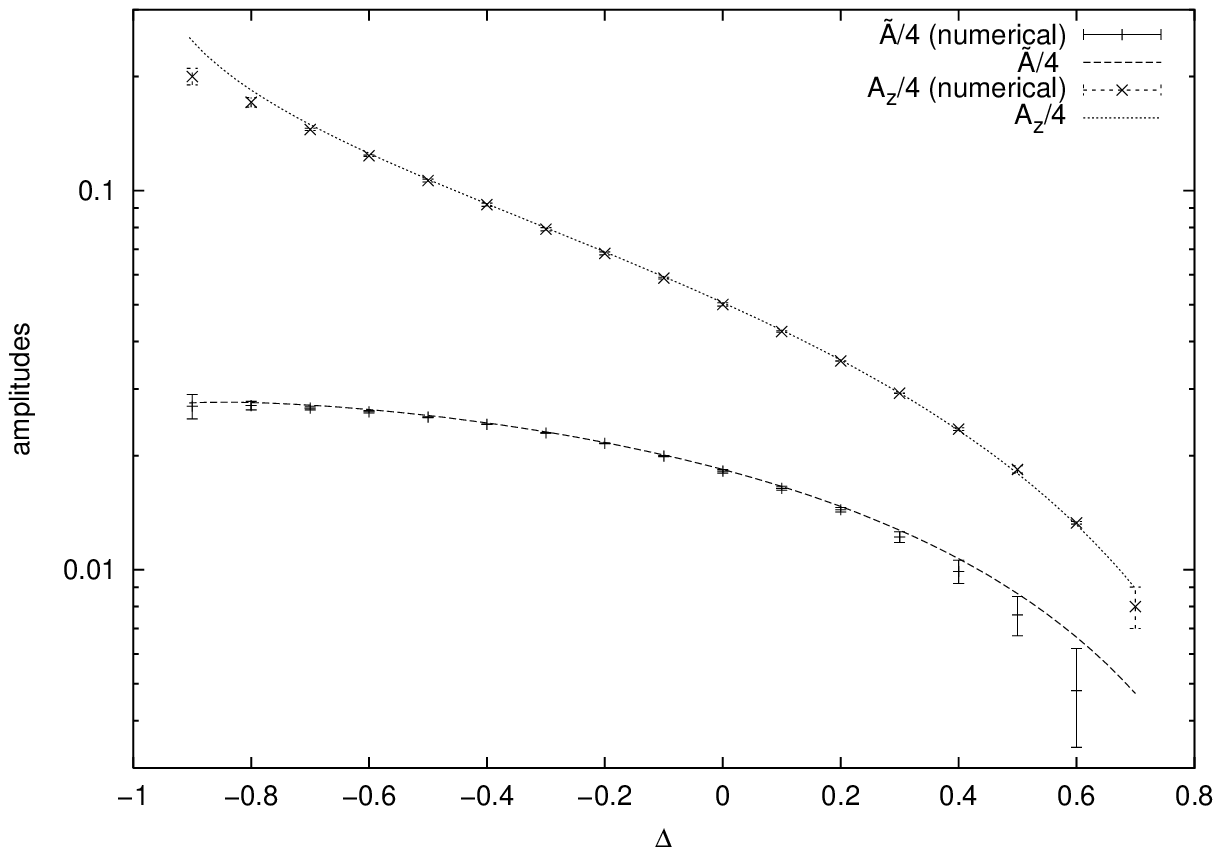}
\caption{The correlation amplitudes $\tilde{A}$ and $A_z$ as functions of
the anisotropy parameter  $\Delta$.
The points with the error bars represent the numerical data obtained
in \cite{HikF98} (see Table~\ref{table-ampl}), 
and the continuous line follows from Eqs.~\eqref{tAx} and \eqref{Az}.}
\label{fig-ampl}
\end{figure}

In \cite{HikF98}, numerical values of the spin-spin equal-time
correlation functions
have been obtained for an open chain of 200 sites by the density-matrix
renormalization-group technique \cite{White}. 
In Table~\ref{table-ampl}, we compare,
for different values of the anisotropy parameter
$\Delta$, the numerical values that follow
from \eqref{tAx},  \eqref{Az} with those estimated in  \cite{HikF98} from the
fitting of the numerical data. The corresponding plots are represented
in Figure~\ref{fig-ampl}.

%
\begin{table}[h]
\begin{center}
\begin{tabular}{| r | l l | l l |}
\hline
\rule{0mm}{4mm} 
$\Delta$\hspace{1mm}  & $\tilde{A}^{NUM}/4$ & 
$\tilde{A}/4$ & $A_z^{NUM}/4$ & $A_z/4$ \\
\hline
\hline
\rule{0mm}{3.6mm} 
0.7  & ***        & 0.004720 & 0.008(1)   & 0.00893 \\
0.6  & 0.0048(14) & 0.006643 & 0.0133(1)  & 0.01314 \\ 
0.5  & 0.0076(9)  & 0.008656 & 0.0184(4)  & 0.01795 \\
0.4  & 0.0099(7)  & 0.010696 & 0.0235(2)  & 0.02332 \\
0.3  & 0.0122(4)  & 0.012717 & 0.02921(3) & 0.02924 \\
0.2  & 0.0144(2)  & 0.014687 & 0.03556(3) & 0.03574 \\
0.1  & 0.0164(2)  & 0.016583 & 0.0425(2)  & 0.04285 \\
0.0  & 0.0182(2)  & 0.018386 & 0.0501(5)  & 0.05066 \\
-0.1 & 0.01995(7) & 0.020082 & 0.0588(3)  & 0.05929 \\
-0.2 & 0.02154(5) & 0.021657 & 0.0683(6)  & 0.06891 \\ 
-0.3 & 0.02296(4) & 0.023098 & 0.0791(8)  & 0.07978 \\
-0.4 & 0.02420(4) & 0.024392 & 0.0918(9)  & 0.09231 \\
-0.5 & 0.02525(6) & 0.025522 & 0.1063(9)  & 0.10713 \\
-0.6 & 0.0261(2)  & 0.026464 & 0.1236(5)  & 0.12539 \\
-0.7 & 0.0267(3)  & 0.027182 & 0.145(1)   & 0.14930 \\
-0.8 & 0.0271(7)  & 0.027608 & 0.171(5)   & 0.18414 \\
-0.9 & 0.027(2)   & 0.027570 & 0.20(1)    & 0.24844 \\ 
\hline
\hline
\end{tabular}
\end{center}
\caption{Correlation amplitudes. $\tilde{A}$ and $A_z$ are computed from
expressions \eqref{tAx}, \eqref{Az}, whereas $\tilde{A}^{NUM}$
and $A_z^{NUM}$ were obtained from the fitting of the numerical data in
\cite{HikF98}. The number in parentheses 
indicate the error bar on the last
quoted digits.}
\label{table-ampl}
\end{table}

\section{Spin-spin correlation functions in the  isotropic limit}
\label{secXXX}

\subsection{Marginal perturbations of the Wess-Zumino-Witten model}

As long as the parameter  $\eta$ is not too close to unity,
the first terms of the  
asymptotic expansions\ \eqref{corr-+}, \eqref{corrzz}
provide a  good
approximation to the spin-spin correlation functions even for
moderate space separations $l$. However, these expansions cannot
be directly applied in the isotropic limit  $\eta\to 1$. 
Indeed, in this limit, the  operator ${\cal O}_{0, 4}+
{\cal O}_{0, -4}$  in the effective Hamiltonian \eqref{Heff}
becomes marginal and induces 
logarithmic corrections  to
the leading power-law asymptotic behavior.

The suitable way to  explore the $\eta\to 1$ limit is
based on  the low-energy
effective theory defined  as a perturbation of the 
Gaussian model with 
$\eta= 1$.
As is well known 
(see e.g.\cite{Coleman}), the  Gaussian model coincides 
in this case with the SU(2) level one Wess-Zumino-Witten 
(WZW) theory.
In particular, the  WZW holomorphic currents 
are identified with the following primary  operators
of the  Gaussian CFT,
\begin{alignat}{2}
  & \J^{z}_{R}=\frac{1}{4}\, 
(\partial_t-\partial_x)\varphi\, ,\qquad & \qquad
  & \J^{\pm}_{ R}={\cal O}_{\pm 1,\mp 2}\, ,\\
  & \J^{z}_{L}=\frac{1}{4}\, (\partial_t+\partial_x)\varphi\, , &
  & \J^{\pm}_{ L}={\cal O}_{\pm 1,\pm 2}\, ,
\end{alignat}
whereas the matrix of the fundamental WZW field is  bosonized as
\begin{equation*}
   \begin{pmatrix}
     {\cal O}_{0,2}     &  i\, {\cal O}_{-1,0}\\
     i\, {\cal O}_{1,0} &  {\cal O}_{0,-2}
   \end{pmatrix}
   \, .
\end{equation*}
The low-energy effective Hamiltonian 
can be expressed as a marginal current-current perturbation of 
the WZW Hamiltonian\ \cite{Aff85},
\begin{equation}
   \H_{\text{XXZ}}=\H_{\text{WZW}}+\int \frac{dx}{2\pi}\, 
               \Big\{\, g_{\parallel}\, \J^{z}_{\! R} \J^{z}_{\! L}
          +\frac{g_{\perp} }{2}\, \big(\, \J_{\! R}^+\J_{\! L}^- 
                                         +\J_{\! R}^-\J_{\! L}^+\,\big)
                          +\cdots\, \Big\}\, .
\end{equation}
In this expression, the coupling constants $g_{\parallel}$ and  $g_{\perp}$
should be understood as ``running'' ones, i.e. depending on
the renormalization scale $r$ which  has the dimension of length.
The corresponding Renormalization Group (RG) 
flow is known as the Kosterlitz-Thouless flow.
For our purpose, we need only to consider the domain
\begin{equation}\label{hsdyusrt}
   |g_{\perp}|\le g_{\parallel}\, ,
\end{equation}
in which all RG
trajectories flow toward the line $g_{\perp}=0$ of the infrared-stable
fixed points associated with the Gaussian CFT.
These trajectories  are  characterized by the limiting values of 
the running coupling $g_{\parallel}$,
\begin{equation}\label{xnchg}
 \epsilon=\frac{1}{2}\, \lim_{r\to +\infty} g_{\parallel}(r)\, ,
\end{equation}
and  the parameter $\epsilon$ 
is simply related with the parameter $\eta$ of the Gaussian model,
\begin{equation}
   \epsilon= 1-\eta\, .
\end{equation}

The RG  flow of the running coupling constants is defined by
a system of  ordinary differential  equations,
\begin{equation}\label{RGexact}
  r\, \frac{dg_{\parallel}}{dr}
    =-\frac{g_{\perp}^2}{f_{\parallel} (g_{\parallel}, g_{\perp})}\ ,
         \qquad\qquad
  r\, \frac{dg_{\perp}}{dr}
    =-\frac{g_{\parallel}\, g_{\perp}} 
{f_{\perp}(g_{\parallel}, g_{\perp})}
\ .
\end{equation}
Perturbatively, the functions $f_{\parallel,\perp}(
g_{\parallel}, g_{\perp})=1+O(g)$
admit loop expansions as power series in $g_{\parallel}$ and
$g_{\perp}$, and  their precise form depends on
the  choice of a renormalization scheme.
We use here the scheme introduced by 
Al.B. Zamolodchikov\cite{Zam95,Alesha},
who showed that, under a suitable diffeomorphism in $g_{\parallel}$
and  $g_{\perp}$, the functions  $f_{\parallel}$ and $f_{\perp}$ 
can be taken equal to each other and to the quantity\footnote{The   
analysis of Refs.\cite{Zam95,Alesha}  concerns  the
ultraviolet-stable domain of the running coupling constants,
which in the current notations is defined as 
$-g_{\parallel}>|g_{\perp}|$. Due to the perturbative nature of
the RG flow equations, the $\beta$-function 
for the infrared-stable  domain\ \eqref{hsdyusrt}\ can 
be obtained through the
formal substitution $g_{\parallel}\to -g_{\parallel}$ in the   original
Zamolodchikov equations.}
\begin{equation}\label{xcn}
   f_{\parallel}=f_{\perp}=1-\frac{g_{\parallel}}{2}\, .
\end{equation}
With this particular choice of the $\beta$-function, 
it is possible to integrate the RG flow equations exactly. 
To do this, let us first note that the system
of differential equations\ \eqref{RGexact}, \eqref{xcn}\ admits a first
integral, the numerical value of which
is determined by means of  the condition\ \eqref{xnchg}:
\begin{equation}\label{xcnlslsk}
   g_{\parallel}^2-g_{\perp}^2=(2 \epsilon)^2\, .
\end{equation}
Then   the equations\ \eqref{RGexact}\ are solved as
\begin{equation}\label{mxncb1}
   g_{\parallel}=2\, \epsilon\  \frac{1+q}{1-q}\, ,\qquad\qquad
   g_{\perp}=4\, \epsilon\  \frac{\sqrt{q}}{1-q}\, ,
\end{equation}
where $q=q(r)$ is the solution of
\begin{equation}\label{mxncb2}
   q^{\frac{1}{2\epsilon}-\frac{1}{2}}\ (1-q)
       =\epsilon\, \Big(\frac{r_0}{r}\Big)^{2}\, . 
\end{equation}
As well as $\epsilon$, 
the dimensional parameter $r_0$ is a RG invariant.
It is of the same order as the lattice  spacing $\varepsilon$,
and is supposed to have a regular loop expansion of the form
\begin{equation}\label{shbcv}
   \frac{\varepsilon}{r_0}= 
   \exp\big(c_0+c_1\epsilon+c_2\epsilon^2+\ldots\, \big)\, .
\end{equation}
It should be noted that the even coefficients $c_0$, $c_2,\ \ldots$
in\ \eqref{shbcv}\ are 
essentially ambiguous and can be chosen as one wants. A variation
of these coefficients corresponds to a smooth redefinition of
the coupling constants which does not affect the $\beta$-function.
On the contrary, the odd constants $c_{2k+1}$ are unambiguous and
precisely specified, once the
form of the RG equations is fixed. It is possible to
show\ \cite{Zam95,Alesha} that the odd constants vanish in
Zamolodchikov's scheme: $c_{2k+1}=0\ (k=0,1\ldots).$
Therefore, once the coefficients $c_{2k}$\ in\ \eqref{shbcv}\ are chosen,
the renormalization scheme is
completely  specified.

As already mentioned, the perturbation by the marginal operators
produces logarithmic corrections 
to  the scale-invariant correlation functions.
Hence, the conformal normalization condition imposed on the
field ${\cal O}_{s,n}$ turns out to be singular for  $\epsilon= 0$, 
and we should   define  renormalized fields 
${\cal O}^{(\text{ren})}_{s,n}$, which 
are  rescaled version of the ``bare'' exponential
operators:
\begin{equation}\label{nsgt}
   {\cal O}^{(\text{ren})}_{s,n}(x,t; r)={\cal Z}^{-\frac{1}{2}}_{s,n}(r)\ 
   {\cal O}_{s,n}(x,t)\,  .
\end{equation}
Notice that, in writing\ \eqref{nsgt}, we assume that there
is no resonance mixing of the operator ${\cal O}_{s,n}$ with other
fields, so that it is renormalized as
a singlet. In particular, one can easily check that this is indeed the case
when
\begin{equation}\label{xbvcpo}
   |n|<2+2\, |s|\, .
\end{equation}
The renormalized fields\ \eqref{nsgt}\ are no longer 
singular at $\epsilon=0$, but depend  on the auxiliary RG scale $r$.
To specify them   completely, we have to impose some  non-singular
normalization condition.
The conventional condition, which   is usually
imposed on Green's  function for a space-like interval
$t^2-x^2<0$, is
\begin{equation}\label{jshdt}
   \big\langle\, T\, {\cal O}^{(\text{ren})}_{s,n}(x,t; r)\,  
   {\cal O}^{(\text{ren})\dagger}_{s,n}(0,0; r)\, \big\rangle
        \big|_{\sqrt{x^2-t^2}=r}=
   \Big(\frac{t-x}{t+x}\Big)^\frac{sn}{2}\, .
\end{equation}
Eqs.~\eqref{nsgt}\ and\ \eqref{jshdt}\ imply that 
the correlators of the ``bare'' exponential  fields should take the
form:
\begin{equation}\label{dkdsui}
   \big\langle\, T\, {\cal O}_{s,n}(x,t)\, {\cal O}^{\dagger}_{s,n}(0,0)\, 
   \big\rangle=
   \Big(\frac{t-x}{t+x}\Big)^\frac{sn}{2}\, 
   {\cal Z}_{ s,n}\big( \sqrt{x^2-t^2}\, \big)\, .
\end{equation}
Then, using\ \eqref{dvt-spin2}, \eqref{dvt-spin1},
one can express the spin-spin correlation functions
through the renormalization factors ${\cal Z}_{ s,n}$. 
For example, for the time-ordered correlation function of $\sigma^x$,
one has: 
\begin{multline}\label{ksjshdyy}
  \langle\,  T\,\sigma_{l+j}^x(t)\, \sigma_{j}^x(0)\,  \rangle\sim
       A\, \varepsilon^{2d_{1,0}}\, 
               {\cal Z}_{ 1,0}\big(\varepsilon \sqrt{l_+l_-}\big)\\ 
   -(-1)^{l}\, {\tilde A}\, \varepsilon^{2d_{1,2}}\, 
          {\cal Z}_{ 1,2}(\varepsilon \sqrt{l_+l_-})\, 
          \bigg\{\, \frac{1}{2}\,
                 \Big(\,\frac{l_+}{l_-}+\frac{l_-}{l_+}\, \Big)
                 - \cR(\varepsilon\sqrt{l_+l_-})\, \bigg\}+\ldots\, .
\end{multline}
Here we use notations\ \eqref{msdshg},
and the function $\cR$ is related to the  following 
causal Green's function as,
\begin{equation}\label{psosi}
   \big\langle\, T\, {\cal O}_{1,2}(x,t)\, {\cal O}_{-1,2}(0,0)\,
       \big\rangle= {\cal Z}_{ 1,2}(r)\, \cR(r)
        \Big|_{r=\sqrt{x^2-t^2}}\, .
\end{equation}

The first terms of the  perturbative expansion for
the scalar factor  ${\cal Z}_{s,n}(r)$ in\ \eqref{nsgt}
can be  deduced from  the results of work\ \cite{Doyon}:
\begin{multline}\label{mnxchyt}
   {\cal Z}_{ s,n}(r)={\bar Z}_{s, n}\,
       \Big(\frac{\varepsilon}{r}\Big)^{\frac{n^2}{4}+s^2 (1+\epsilon^2)}\ 
       \big(g_{\perp}^2\big)^{\frac{n^2}{16}-\frac{s^2}{4}(1-\epsilon^2)}\\ 
   \times
   e^{u_1g_{\parallel}+u_2g_{\parallel}^3}\  
       \Big(1+g_{\perp}^2\, (v_1-v_2\, g_{\parallel})+O(g^4)\, \Big)\, ,
\end{multline}
where
\begin{equation*}
   {\bar Z}_{s, n}=\varepsilon^{-2d_{s,n}}\
   \Big(\, 2^{1-\epsilon}\, \sqrt{\epsilon}\ 
           e^{-c_0\epsilon-c_2\epsilon^3+\ldots}
   \Big)^{2s^2-2d_{s,n}}\ 
   e^{-2\, \epsilon \, u_1-(2 \epsilon)^3 \, u_2+\ldots}\, ,
\end{equation*}
and $d_{s,n}$ is given by\ \eqref{dimen}\ with $\eta=1-\epsilon$.
The coefficients $u_1$, $u_2$, $v_1$ and $v_2$ in
these equations are listed in Appendix C. Notice that
to derive\ \eqref{mnxchyt}\ one should assume that the field
$\cO_{s,n}$ is mutually local with respect to the
density of the effective Hamiltonian\ \eqref{Heff}. This
assumption implies that $s\in {\mathbb Z}$, but does not impose  any
restriction on
$n$ in addition to\ \eqref{xbvcpo}.

As follows from the Callan-Symanzik equation, the
function $\cR$\ in\ \eqref{psosi}\ admits 
a perturbative expansion in terms
of  the running coupling constants.  Explicitly, one can  obtain
\begin{equation}\label{hsgdf}
\cR=-\frac{g_{\perp}}{4}\ \bigg\{\, 
g_{\parallel}+\Big(\, c-\frac{1}{2}\, \Big)\, 
g^2_{\parallel}+c\, g^2_{\perp}+O(g^3)\, \bigg\}\ .
\end{equation}
The constant $c$  appearing  in Eq.~\eqref{hsgdf} is
related to  $c_0$\ from\ \eqref{shbcv} as
\begin{equation}\label{bxvcf}
    c_0=c+\gamma_E+\frac{1}{2}\, \ln(2\pi)\, ,
\end{equation}
where $\gamma_E=0.5772\ldots$ is the Euler constant.
Combining relation\ \eqref{ksjshdyy}\
with \ \eqref{mnxchyt}\ and\ \eqref{hsgdf},
one can deduce  the RG improved  expansion of  the $\sigma^x$ lattice
correlator
which is  applicable for $\epsilon\ll 1$.

We can similarly derive an expansion
for  the correlation function of $\sigma^z$. 
The  relation 
\begin{equation}\label{ansju}
   \frac{1}{2\pi \sqrt{\eta}}\, \partial_t\varphi =\frac{2}{i\pi}\,
       \partial_x\partial_n{\cal O}_{0,n}\Big|_{n=0}\, ,
\end{equation}
which follows from the definition\ \eqref{op},
is useful to perform this computation.
Note also that the operators
\begin{equation*}
   {\cal O_+}=\frac{{\cal O}_{0,2}+{\cal O}_{0,-2}}{\sqrt{2}}\, ,
           \qquad\qquad
   {\cal O_-}=\frac{{\cal O}_{0,2}-{\cal O}_{0,-2}}{\sqrt{2}\, i}\, ,
\end{equation*}
renormalize as singlets:
\begin{equation}\label{laksiu}
   {\cal O}^{(\text{ren})}_{\pm}(x,t;r)=
     {\cal Z}^{-\frac{1}{2}}_{\pm}(r)\ {\cal O}_{\pm}(x,t)\, .
\end{equation}
Indeed, since
\begin{equation*}
   {\mathbb C}\, {\cal O_\pm } \, {\mathbb C}=\pm {\cal O}_\pm \, ,
\end{equation*}
invariance with respect to the charge conjugation
prevents  
resonance mixing of ${\cal O}_+$ and  ${\cal O}_-$. 
Now, using equations\ \eqref{ansju}, \eqref{laksiu} 
and\ \eqref{dvt-spin1}, one obtains
\begin{multline}\label{kassdyy}
   \langle\,  T\, \sigma_{l+j}^z(t)\, \sigma_{j}^z(0)\,  \rangle\sim
        -\frac{2}{\pi^2}\, 
          (\partial_{l_+}+\partial_{l_-})^2\partial_n^2\, 
          {\cal Z}_{0,n}\big(\varepsilon\sqrt{l_+l_-}\big)\Big|_{n=0}\\ 
   +(-1)^l\, A_z\, \varepsilon^{2d_{0,2}}\, 
                {\cal Z}_-(\varepsilon\sqrt{l_+l_-})+\ldots\, .
\end{multline}

We collect in Appendix D the  RG improved expansions
of the different-time  two-point correlation functions
\eqref{ksjshdyy}\ and\ \eqref{kassdyy}.

\subsection{Equal-time  correlation functions for the XXX spin chain.
Comparison with numerical results}

Using expansions\ \eqref{bcfr}, \eqref{bcfr1}, it
is easy to perform the isotropic limit.
Setting $\epsilon=0$ and 
$g_{\perp}=g_{\parallel}=g$, one obtains the following
large $l$ expansion
for the equal-time spin-spin correlation functions\footnote{The 
coefficient $\sqrt{2/\pi^3}$ in Eq.~\eqref{W(n)} was originally
obtained in Ref.~\cite{aff}.}:
\begin{align}
   \bracket{  \sigma^x_{l+j}\, \sigma_j^{x}  } 
   &= (-1)^l\ \bracket{  \sigma^z_{l+j}\, \sigma_j^{z} }\nonumber\\
   & \sim \sqrt{\frac{2}{\pi^3}}\, \frac{1}{l\,  \sqrt{g}}\, 
   \bigg\{\,  1+\Big(\, \frac{3}{8}-\frac{c}{2}\,\Big)\, g
          +\Big(\,\frac{5}{128}-\frac{c}{16}
          -\frac{c^2}{8}\,\Big)\, g^2\nonumber\\
   &\ \qquad\       +\Big(\,\frac{21}{1024}+\frac{7 c}{256}
          -\frac{7 c^2}{64}-\frac{c^3}{16}
          +\frac{13\, \zeta(3)}{32}\,\Big)\, g^3+O(g^4)\, \bigg\}\nonumber\\
   &\ \qquad\      -\frac{(-1)^l}{\pi^2\,  l^2}\,  
          \bigg\{\, 1+\frac{g}{2}+\Big(c+\frac{3}{4}\,\Big)\,
           \frac{g^2}{2}+\frac{c(c+2)}{2}\, g^3+O(g^4) \, \bigg\}+\ldots\ .
   \label{W(n)}    
\end{align}
Here $g=g(l)$ is a solution of the equation
\begin{equation}\label{cju}
   \sqrt{g}\, e^\frac{1}{g}
      = 2\sqrt{2\pi}\, e^{\gamma_E+c}\,  l \, , 
\end{equation}
which corresponds to the limit  $\epsilon\to 0$ of 
Eqs.~\eqref{mxncb1} and \eqref{mxncb2}.
Let us stress that, if the perturbation series in\ \eqref{W(n)}\
can be summed,
then the  correlation function should not
depend on the auxiliary parameter $c$\ \eqref{shbcv}, \eqref{bxvcf}. 
This, however, is not true
if we truncate the perturbative series at some finite order.
Thus, when fitting the numerical 
data with\ \eqref{W(n)}, we may treat $c$ as
an optimization parameter, allowing us to minimize the 
remainder of the series, or at least to develop 
some feeling concerning the effects of this remainder. 

\begin{figure}
\centering
\includegraphics[width=12cm]{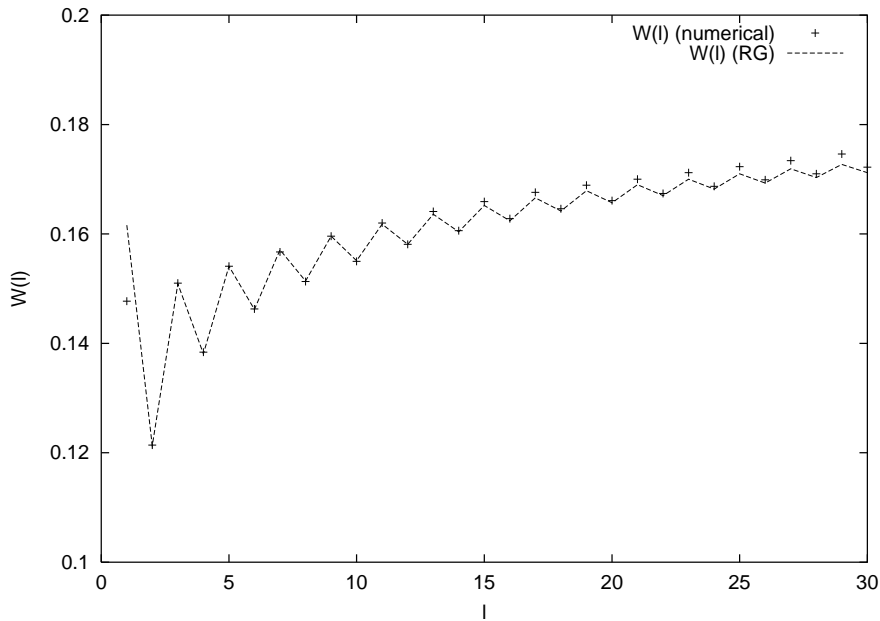}
\caption{The correlation function 
$W(l)=\frac{l}{4}\ \bracket{  
\sigma^x_{l+j}\,  \sigma^x_{j}}$ of the XXX spin chain 
according to the distance $l$. 
The dots represent the numerical data obtained
in \cite{HalHM95}, and the continuous 
line connects
odd and even terms obtained from  Eq.~\eqref{W(n)}\ with $c=-1$.} 
\label{XX1}
\end{figure}

The correlation function \eqref{W(n)} has been computed numerically in
\cite{HalHM95} for $1\le l\le 30$. 
The authors used the density-matrix algorithm \cite{White}
to study the
large-distance decay of the correlation function for XXX spin chains
with  $14\leq N\leq 70$ sites. To extract the values of the
correlation function in the infinite chain case, they adopted
the phenomenological scaling relation of Kaplan {\em et al.}\cite{Kaplan} 
(see also\ \cite{AfB}).
The relative error of the interpolation procedure was estimated 
to be of order $1\%$ 
for the largest $l$ values.
In Table \ref{table2}, we compare those numerical data to the
results obtained from \eqref{W(n)} in the cases $c=-1$ and $c=-2$.
The corresponding plots (numerical data against RG result for $c=-1$)
are given in Figure \ref{XX1}.
It appears that  the numerical data are
consistent with our prediction within the given errors.
\begin{table}[h]
\begin{center}
\begin{tabular}{| c c c c || c c c c |}
\hhline{----||----}

$l$  & NUM  & RG\,($c=-1$) & RG\,($c=-2$) & $l$  & NUM  & RG\,($c=-1)$ 
& RG\,($c=-2$) \\
\hhline{====::====}
1  & 0.1477  & 0.1616 & ***    & 16 & 0.1628  & 0.1624 & 0.1630 \\
2  & 0.1214  & 0.1213 & 0.1918 & 17 & 0.1676  & 0.1666 & 0.1671 \\
3  & 0.1510  & 0.1509 & 0.1583 & 18 & 0.1646  & 0.1642 & 0.1647 \\
4  & 0.1384  & 0.1381 & 0.1424 & 19 & 0.1689  & 0.1679 & 0.1683 \\
5  & 0.1541  & 0.1541 & 0.1566 & 20 & 0.1661  & 0.1657 & 0.1661 \\
6  & 0.1463  & 0.1462 & 0.1482 & 21 & 0.1700  & 0.1690 & 0.1694 \\
7  & 0.1567  & 0.1571 & 0.1586 & 22 & 0.1674  & 0.1670 & 0.1674 \\
8  & 0.1513  & 0.1514 & 0.1527 & 23 & 0.1712  & 0.1700 & 0.1704 \\
9  & 0.1596  & 0.1596 & 0.1607 & 24 & 0.1687  & 0.1682 & 0.1686 \\
10 & 0.1550  & 0.1552 & 0.1561 & 25 & 0.1723  & 0.1710 & 0.1713 \\
11 & 0.1620  & 0.1618 & 0.1626 & 26 & 0.1699  & 0.1693 & 0.1697 \\
12 & 0.1581  & 0.1581 & 0.1588 & 27 & 0.1734  & 0.1719 & 0.1722 \\
13 & 0.1641  & 0.1636 & 0.1643 & 28 & 0.1710  & 0.1703 & 0.1706 \\
14 & 0.1606  & 0.1604 & 0.1611 & 29 &  0.1746 & 0.1727 & 0.1730 \\
15 & 0.1659  & 0.1652 & 0.1658 & 30 & 0.1722  & 0.1712 & 0.1716 \\
\hhline{====:b:====}
\end{tabular}
\end{center}
\caption{Numerical values of the correlation function 
$\frac{l}{4}\, \bracket{ \sigma^x_{l+j}\, \sigma^x_{j} }$ 
of the XXX spin chain 
according to the distance $l$. The column ``NUM'' has been obtained
in \cite{HalHM95}, whereas the columns ``RG\,($c=-1$)'' and ``RG\,($c=-2$)'' 
follow from \eqref{W(n)} with the corresponding values of the
free parameter $c$.}
\label{table2}
\end{table}

\subsection{Erratum of\ \cite{Luk98} }

The spin-spin correlation function
in the limit $\epsilon\to 0$
was previously studied in Section 5 of  Ref.~\cite{Luk98}.
The analysis performed in that paper, along
with the numerical results obtained in \cite{HalHM95} (see
Fig.~2 of \cite{Luk98}), strongly suggested the existence of
an additional  staggered term of the
form
\begin{equation*}
   \propto \frac{(-1)^l+1}{l^{\eta+1}} 
\end{equation*}
in the large\ distance asymptotic expansion  of
the correlation function\ $\langle\, \sigma_{l+j}^x\,
 \sigma_{j}^x\, \rangle$.
It was argued in \cite{Luk98} that
such  a term occurs because of
the presence  of  a correction
of 
the type
$ (-1)^l\ \partial_x {\cal O}_{\pm 1,0}$
in expansions \eqref{dvt-spin2}. 
However, the RG computation from \cite{Luk98} appears to be
erroneous. Indeed, Eqs.~(5.6) and (5.7) from\ \cite{Luk98}\ have 
to be replaced respectively by our equations
\eqref{bcfr} and \eqref{W(n)}.
Therefore, contrary to what was claimed in \cite{Luk98},
the numerical data are  consistent  (within the numerical errors)
with\  \eqref{dvt-spin2}.

\section{Conclusion and further remarks}

The purpose of this  article is the quantitative study of the  long-distance
behavior of spin-spin correlation functions for the
XXZ Heisenberg chain in the critical regime.
Our main result here is the determination of analytical expressions
for the correlation amplitudes 
involved in the corresponding asymptotic expansions.
To obtain these values, we considered quantum field theory
which describes the scaling limit of the lattice model,
and compared, in this limit,
the respective normalizations 
of the lattice operators and of the  corresponding local fields.
This comparison was achieved by considering known  exact 
matrix elements 
(form factors).

\bigskip
\noindent
We would like to  conclude the article with the following remarks.

$\bullet$ The method used in this work    can  be 
applied to higher order terms in expansion 
\eqref{dvtz} or \eqref{dvt+-}.
For example, we were able to
compute (up to  sign factors)
all  constants $C_m$ from\ \eqref{dvt+-}\ for  odd integers $m=2p+1$:
\begin{multline}\label{conjCm}  
    (C_{2p+1})^2=\frac{2}{\eta (1-\eta)}\, 
      \bigg[\frac{ \Gamma(\frac{\eta}{2-2\eta})}
                 {2\sqrt{\pi}\, \Gamma(\frac{1}{2-2\eta})}
      \bigg]^{\eta+\frac{(2p+1)^2}{\eta}}
      \prod_{j=1}^{p}
             \bigg\{\sin^2\Big(\frac{2\pi j}{\eta}\Big)\,
                   \cot^2\Big(\frac{\pi  (2j-1)}{2-2\eta}\Big)\bigg\}\\
   \qquad\qquad\times \exp\bigg\{-\int_0^{\infty}\frac{dt}{t}\, 
      \Big(\frac{\cosh(2 \eta t) e^{-2 (2p+1) t}-1}
                {2\sinh(\eta t)\sinh(t)\cosh((1-\eta)t)}\\
   +\frac{2p+1}{\sinh(\eta t)}-\Big(\eta+\frac{(2p+1)^2}{\eta}
\,\Big)\, e^{-2 t}\, \Big)\, \bigg\}\ .
\end{multline}

\bigskip

$\bullet$ 
One can also study  expansions of   other local 
lattice operators in terms of the  scaling fields. 
For example,  we have calculated the  constant $C_0^{(s)}$ in 
the leading term of  the expansion of the lattice operators
\begin{equation}
   \sg^\pm_l \sg^\pm_{l+1}\dots\sg^\pm_{l+s-1}\sim C^{(s)}_0\,
                      \eps^{\frac{s^2\eta}{2}}\, \cO_{\pm s,0}+\ldots\, ,
\end{equation} 
for which 
we obtained $(p=0,1\ldots)$:
\begin{align}\label{F2p}
   &C^{(2p)}_0=\bigg[\frac{\Gamma(\frac{\eta}{2-2\eta})}
                       {2\sqrt{\pi}\, \Gamma(\frac{1}{2-2\eta})}
                                           \bigg]^{2\eta p^2}
            \frac{1}{\pi^{p}(1-\eta)^{2p^2-p}} 
          \prod_{j=1}^{p}
          \frac{\Gamma^2(\frac{1}{2}+\frac{\eta(2j-1)}{2-2\eta})\, 
                         \Gamma^2( \eta(2j-1))}
               {\Gamma^2(\frac{\eta(2j-1)}{2-2\eta})}\ ,\\
   &C^{(2p+1)}_0= \bigg[\frac{\Gamma(\frac{\eta}{2-2\eta})}
                       {2\sqrt{\pi}\, \Gamma(\frac{1}{2-2\eta})}
                                           \bigg]^{2\eta (p+\frac{1}{2})^2}
            \frac{1}{2\pi^{p}(1-\eta)^{2p^2+p+1}} 
          \prod_{j=1}^{p}
          \frac{\Gamma^2(\frac{1}{2}+\frac{\eta j}{1-\eta})\, 
                         \Gamma^2( 2\eta j)}
               {\Gamma^2(\frac{\eta j}{1-\eta})}\nonumber\\
   &\hspace{3.4cm} \times
          \exp\bigg\{-\int_{0}^{\infty}\frac{dt}{t}\ 
              \Big(\, \frac{\sinh(\eta t)}{2\sinh(t)\cosh((1-\eta)t)}
                         -\frac{\eta}{2}\ e^{-2t}\, \Big)\, \bigg\}\ .
   \label{F2p+1}
\end{align}

$\bullet$ 
Eventually, one can wonder if it is possible to confirm
our predictions
from existing integral  representations of lattice correlators.
Up to now, although 
explicit expressions for the equal-time 
spin-spin correlation functions at finite lattice 
distances are known\ \cite{JimMbo,KMST}, 
their long-distance behavior
was  studied only  for the so-called ``free fermion point'',
$\Delta=0$. In this case the XXZ spin chain
can be mapped onto
two non-interacting critical Ising models and the long-distance
asymptotics are readily derived from   results of 
works\ \cite{LieSM61,McCoyT,Perk}.

$\bullet$ The approach of this work
is actually quite general for lattice solvable models at criticality:
from a  knowledge of particular form factors
of a lattice theory and of its quantum field theory counterpart at the
scaling limit, it is possible to predict the   amplitudes
which govern the large distance behavior of  lattice correlation
functions. It would indeed be interesting to obtain effective results for
other critical exactly solvable lattice models.

\section*{Acknowledgments}

\bigskip
We  would like to thank A. Furusaki for kindly providing us
with the numerical
data of work \cite{HikF98} and  A.B. Zamolodchikov for 
interesting discussions.
S.L.  acknowledge helpful discussions with  V.V. Bazhanov.
We are also grateful to Daniela Kusmierek for her careful reading of the
manuscript. 

\bigskip
\noindent
This research is supported in part by DOE grant
$\#$DE-FG02-96ER40959. V.T. is also supported in part  by CNRS.

\appendix

\section{Two-particle form factors in the  XYZ model}\label{ap-XYZ}

In this appendix, we collect  explicit expressions of 
two-particle form factors 
of local spin operators in the XYZ model.
Following Baxter\ \cite{Baxter2,Bax82L},  we use the  parameterization
of the coupling constants  $J_x > J_y> |J_z|$ of the
Hamiltonian\ \eqref{HXYZ} in terms of $0<\eta<1$ and of the
elliptic nome $0<p< 1$:
\begin{align}\label{sdla1}
   &J_x= \frac{1-\eta}{\pi\eps}\ \biggl( 
           \frac{\vth_4(\eta) \vth'_1(0)}{\vth_4(0)\vth_1(\eta)}
       + \frac{\vth_1(\eta) \vth'_1(0)}{\vth_4(0)\vth_4(\eta)} \biggr)\, ,\\
   &J_y= \frac{1-\eta}{\pi\eps}\ \biggl( 
        \frac{\vth_4(\eta) \vth'_1(0)}{\vth_4(0)\vth_1(\eta)}
       - \frac{\vth_1(\eta) \vth'_1(0)}{\vth_4(0)\vth_4(\eta)} \biggr)\, ,\\
  \label{sdla3}
  &J_z= \frac{1-\eta}{\pi\eps}\  \biggl( 
           \frac{ \vth'_1(\eta)}{\vth_1(\eta)}
           - \frac{\vth'_4(\eta)}{\vth_4(\eta)} \biggr)\, .
\end{align}
Here $\vth_i(u)\equiv \vth_i(u,p)$ 
denote the elliptic theta-functions 
\begin{align}
   &\vth_1(u, p)=2 p^{1/4}\  \sin(\pi u)
\ \prod_{n=1}^\infty 
                 (1-p^{2n})(1-2p^{2n} \cos (2\pi u)+ p^{4n})\, ,\\
   &\vth_4(u, p)=\prod_{n=1}^\infty 
                 (1-p^{2n})(1-2p^{2n-1} \cos (2\pi u)+ p^{2(2n-1)})\, ,
\end{align}
and the prime in Eqs.~\eqref{sdla1}-\eqref{sdla3}\
means a derivative: $\vth'_1=\partial_u\vth_1$. We shall also
use the other conventional theta-functions
\begin{equation*}
   \vth_2(u)=\vth_1(u+1/2)\, ,\qquad \qquad\  
   \vth_3(u)=\vth_4(u+1/2)\, ,
\end{equation*}
and the notation,
\begin{equation}\label{msdfh}
   \xi=\frac{\eta}{1-\eta}\ .
\end{equation}

With this  parameterization,
the VEV of $\sg^x$\ \eqref{vxcd}\ is  given 
by the Baxter-Kelland formula~\cite{BaxK74}:
%
%
\begin{equation}\label{B-K}
         F=(1+\xi)\  p^\frac{\xi}{8}\ 
             \pl_{n=1}^{\infty}
        \biggl(\, \frac{1-p^{n(1+\xi)}}{1-p^{(n-\frac{1}{2})(1+\xi)} }\ 
                \frac{1-p^{n-\frac{1}{2}}}{1-p^{n}}\, \biggr)^2\  . 
\end{equation}

In order to describe the two-particle form factors, 
one needs to know the explicit form
of the  dispersion relation\ \eqref{transaa}.
For this purpose, it is convenient to  parameterize
the quasi-momentum $k$ by means of  the  so-called rapidity variable 
$\theta$:
\begin{equation}\label{poi}
  e^{ik(\theta)}=
  \frac{\vartheta_4\big(\, \frac{\theta}{2 i\pi}-\frac{1}{4}\, ,\, 
        p^{\frac{1+\xi}{4}}\, \big)}
       {\vartheta_4\big(\,\frac{\theta}{2 i\pi}+\frac{1}{4}\, ,\, 
        p^{\frac{1+\xi}{4}}\,  \big)}\, .
\end{equation}
As a function of $\theta$, the excitation energy explicitly reads 
\cite{JohKM73},
\begin{equation}\label{poii}
   {\cal E}(\theta)=\frac{\partial k(\theta)}{\partial \theta}\, .
\end{equation}
Equations\ \eqref{poi}\ and\ \eqref{poii}\ define 
the dispersion relation ${\cal E} ={\cal E}(k)$ in 
parametric form.

The two-particle form factors of the spin operators can be  computed
by means of the $q$-vertex operator approach, 
for which progress has been made recently in the XYZ case
\cite{LasP98,Las01}. In \cite{Las01},  the two-particle form factors of the 
$\sg^x$ operator were obtained%
\footnote{%
Note that the regime considered in \cite{Las01} is the so-called principal 
one ($-J_z > J_x\geq |J_y|$). 
To apply  the  results obtained there to the
case with 
$J_x > J_y\geq |J_z|$, one has to replace $\sg_l^x$, $\sg_l^y$ and 
$\sg_l^z$ from\ \cite{Las01}\ 
respectively by $\sg_l^y$, $(-1)^l\sg_l^z$ and $(-1)^l\sg_l^x$, which 
corresponds to a similarity transformation of the Hamiltonian
\eqref{HXYZ}.
}:
\begin{align}\label{ksiy22}
&F_1^x=  
       \frac{ F_0\ {\bar G}(\theta_1-\theta_2\, ,\, p )\
     \vartheta_4\big(\, 0\, ,\, p^\frac{1}{2}\, \big)}
            {\vartheta_4\big(\, \frac{\theta_1}{2 i\pi}-\frac{1}{4}\, ,\, 
  p^{\frac{1+\xi}{4}}\, \big)\                               
             \vartheta_4\big(\, \frac{\theta_2}{2 i\pi}-\frac{1}{4}\, ,\, 
  p^{\frac{1+\xi}{4}}\, \big)}\ \ 
       \frac{\vartheta_4\big(\, \frac{\theta_1+\theta_2}{2 i\pi}\, ,\,
                                      p^{\frac{1+\xi}{2}}\, \big)}
            {\vartheta_1\big(\, \frac{\theta_1-\theta_2+i\pi}{2 i\pi\xi }\, ,
                             \,  p^{\frac{1+\xi}{2\xi}}\, \big)}\, , \\ 
\label{ksiy22a}
&F_2^x=\frac{ F_0\ {\bar G}(\theta_1-\theta_2\, ,\, p )\
     \vartheta_4\big(\, 0\, ,\, p^\frac{1}{2}\, \big)}
            {\vartheta_4\big(\, \frac{\theta_1}{2 i\pi}-\frac{1}{4}\, ,\,
  p^{\frac{1+\xi}{4}}\, \big)\
             \vartheta_4\big(\, \frac{\theta_2}{2 i\pi}-\frac{1}{4}\, ,\,
  p^{\frac{1+\xi}{4}}\, \big)}\ \
       \frac{\vartheta_1\big(\, \frac{\theta_1+\theta_2}{2 i\pi}\, ,\,
                                      p^{\frac{1+\xi}{2}}\, \big)}
            {\vartheta_4\big(\, \frac{\theta_1-\theta_2+i\pi}{2 i\pi\xi }\, ,
                             \,  p^{\frac{1+\xi}{2\xi}}\, \big)}\, .
\end{align}
For $-2\pi<\Im m(\theta)<0$ the  
meromorphic function ${\bar G}$ reads
\begin{equation}\label{oslmx}
   \bar{G}(\theta ,\, p)
        = e^{\frac{\delta(1+\xi)}{8\pi\xi}\,
 (\theta+i\pi)^2}\, 
          \exp\bigg\{\, \sum_{n=1}^{\infty}\, \frac{1}{n}\
          \frac{\sin^2(\delta n (\theta+i\pi)/2)\, 
                \sinh(\pi\delta(\xi+1) n/2)}
               {\sinh(\pi\delta n)\, \sinh(\pi\delta\xi n/2)\,
                \cosh(\pi \delta n/2) }\, \bigg\}\, ,
\end{equation}
and it  is defined through an analytic continuation
outside this domain.
The parameter $\delta$ in \eqref{oslmx}\  is 
related to the elliptic nome as
$p=e^{-\frac{4\pi}{\delta(\xi+1)}},$ and
the constant $F_0$ in\ \eqref{ksiy22}, \eqref{ksiy22a}\ is given by 
\begin{equation*}
   F_0=\frac{1+\xi}{\pi\xi }\ \ 
       \frac{\theta'_1\big(\,0\, ,\, p^{\frac{1+\xi}{2}}\, \big)\ 
             \theta'_1\big(\,0\, ,\, p^{\frac{1+\xi}{2\xi}}\, \big)}
            {\theta'_1\big(\,0\, ,\, p^\frac{1}{2}\,\big)}\, .
\end{equation*}
Notice that, in writing the form factors, we always
assume the conventional normalization of   vacuum states,
$\langle\, {\rm vac}\, |\,  {\rm vac}\, \rangle=1$, and of
in-asymptotic states:
\begin{equation}\label{mxtr}
   {}_{\rm in}\big\langle\, {\bf  B}_{\sigma'_{n}}(k'_{n})\ldots
        {\bf B}_{\sigma'_{1}}(k'_{1})\,
        |\, {\bf B}_{\sigma_{1}}(k_{1})\ldots {\bf B}_{\sigma_{n}}(k_{n})\,
        \big\rangle_{\rm in}
   =(2\pi)^n\ \prod_{j=1}^n\delta_{\sigma_j \sigma'_j}\
        \delta(\theta_j-\theta'_j)\, ,
\end{equation}
where $k_j=k(\theta_j)$ and $k'_j=k(\theta'_j)$.

Using the method proposed in \cite{Las01}, one can also compute the 
two-particle form factors of the other spin fields, $\sigma^y$ and
$\sigma^z$:
\begin{align}\label{ff22}
&F_1^y=-\ 
       \frac{ F_0\ {\bar G}(\theta_1-\theta_2\, ,\, p )\
     \vartheta_3\big(\, 0\, ,\, p^\frac{1}{2}\, \big)}
            {\vartheta_4\big(\, \frac{\theta_1}{2 i\pi}-\frac{1}{4}\, ,\,
  p^{\frac{1+\xi}{4}}\, \big)\
             \vartheta_4\big(\, \frac{\theta_2}{2 i\pi}-\frac{1}{4}\, ,\,
p^{\frac{1+\xi}{4}}\, \big)}\ \
       \frac{\vartheta_3\big(\, \frac{\theta_1+\theta_2}{2 i\pi}\, ,\,
                                      p^{\frac{1+\xi}{2}}\, \big)}
            {\vartheta_2\big(\, \frac{\theta_1-\theta_2+i\pi}{2 i\pi\xi }\, ,
                             \,  p^{\frac{1+\xi}{2\xi}}\, \big)}\, , \\
&F_2^y=-\ 
    \frac{ F_0\ {\bar G}(\theta_1-\theta_2\, ,\, p )\
     \vartheta_3\big(\, 0\, ,\, p^\frac{1}{2}\, \big)}
            {\vartheta_4\big(\, \frac{\theta_1}{2 i\pi}-\frac{1}{4}\, ,\,
  p^{\frac{1+\xi}{4}}\, \big)\
             \vartheta_4\big(\, \frac{\theta_2}{2 i\pi}-\frac{1}{4}\, ,\,
p^{\frac{1+\xi}{4}}\, \big)}\ \
       \frac{\vartheta_2\big(\, \frac{\theta_1+\theta_2}{2 i\pi}\, ,\,
                                      p^{\frac{1+\xi}{2}}\, \big)}
            {\vartheta_3\big(\, \frac{\theta_1-\theta_2+i\pi}{2 i\pi\xi }\, ,
                             \,  p^{\frac{1+\xi}{2\xi}}\, \big)}\, .
\end{align}
and
\begin{align}\label{ffz}
     &F_1^z =i\  \ \frac{ F_0\ {\bar G}(\theta_1-\theta_2\, ,\, p )\
     \vartheta_2\big(\, 0\, ,\, p^\frac{1}{2}\, \big)}
            {\vartheta_4\big(\, \frac{\theta_1}{2 i\pi}-\frac{1}{4}\, ,\,
  p^{\frac{1+\xi}{4}}\, \big)\
             \vartheta_4\big(\, \frac{\theta_2}{2 i\pi}-\frac{1}{4}\, ,\,
p^{\frac{1+\xi}{4}}\, \big)}\ \
       \frac{\vartheta_2\big(\, \frac{\theta_1+\theta_2}{2 i\pi}\, ,\,
                                      p^{\frac{1+\xi}{2}}\, \big)}
            {\vartheta_2\big(\, \frac{\theta_1-\theta_2+i\pi}{2 i\pi\xi }\, ,
                             \,  p^{\frac{1+\xi}{2\xi}}\, \big)}\, ,
\\
&F^z_2=i\  \ \frac{ F_0\ {\bar G}(\theta_1-\theta_2\, ,\, p )\
     \vartheta_2\big(\, 0\, ,\, p^\frac{1}{2}\, \big)}
            {\vartheta_4\big(\, \frac{\theta_1}{2 i\pi}-\frac{1}{4}\, ,\,
  p^{\frac{1+\xi}{4}}\, \big)\
             \vartheta_4\big(\, \frac{\theta_2}{2 i\pi}-\frac{1}{4}\, ,\,
p^{\frac{1+\xi}{4}}\, \big)}\ \
       \frac{\vartheta_3\big(\, \frac{\theta_1+\theta_2}{2 i\pi}\, ,\,
                                      p^{\frac{1+\xi}{2}}\, \big)}
            {\vartheta_3\big(\, \frac{\theta_1-\theta_2+i\pi}{2 i\pi\xi }\, ,
                             \,  p^{\frac{1+\xi}{2\xi}}\, \big)}\, .
\end{align}

To compare the lattice and the sine-Gordon two-particle form factors,
one should take the
limit $p\tend 0$. Notice that
\begin{equation*} 
   p \simeq  (4\, R_c)^{-\frac{4}{\xi+1}}\, ,
\end{equation*}
where the correlation length is defined as in\ \eqref{nxbt}. 
The following relation between the function
${\bar G}$\ \eqref{oslmx}\ and 
the sine-Gordon minimal form factor\ \eqref{G}:  
\begin{equation*}
   \lim_{p\to 0} {\bar G}(\theta\, ,\, p)=\frac{G(\theta)}{G(-i\pi)}\, .
\end{equation*}
is useful to proceed with this limit.

\section{Form factors of topologically charged operators in the
                      sine-Gordon model}\label{ap-SG}

In this appendix, we recall the expressions obtained in \cite{LukZ01}
concerning the form factors of 
the topologically charged (or soliton-creating)
operators $\cO_{s, n}$ in the sine-Gordon model.
The simplest non-vanishing  form factor of 
$\cO_{s,n}$ is given by the 
formula:
\begin{equation}
   \big\langle\, {\cO}_{s, n}(0)\,|\,
    {\bf A}_{-}(\theta_1)\cdots {\bf A}_{-}(\theta_n)\big\rangle_{\rm in}
       =\sqrt{{ Z}_{s,n}}\ e^{\frac{i{\pi n s}}{4}}\,
        \prod_{m=1}^{n}\,e^{\frac{s \theta_m}{2}}\,
        \prod_{m<j}\,G(\theta_m - \theta_j)\, .
\end{equation}
Here, the minimal form factor $G$ has a form:
\begin{equation}
   G(\theta)=i\, {\cC}_1\, \sinh(\theta/2)\ 
                 \exp \bigg\{\int_0^{\infty}\frac{dt}{t}\
                 \frac{\sinh^2(t(1-i\theta/\pi))\,\sinh((\xi-1) t)}
                      {\sinh(2t)\, \cosh(t)\,  \sinh(\xi t)}\, \bigg\}\ .
          \label{G}
\end{equation}
The explicit expression of the normalization constant
${ Z}_{s,n}$, which has been conjectured in \cite{LukZ01},
is the following:
\begin{align}
   &{ Z}_{s,n}=   \Big(\frac{{\cC}_2}{2\, \cC_1^2}\Big)^{\frac{n}{2}}\
                       \Big(\frac{\xi\, {\cC}_2}{16} \Big)^{-\frac{n^2}{4}}\
       \bigg[\, \frac{\sqrt{\pi} M \Gamma\big(\frac{3}{2}+\frac{\xi}{2}\big)}
                    {\Gamma\big(\frac{\xi}{2}\big)}\, \bigg]^{2d_{s,n}}
              \nonumber\\
   &\ \times
       \exp\bigg\{ \int_{0}^{\infty} \frac{dt}{t}\,
         \Big[\, \frac{\cosh(2\xi s t)\, e^{-(1+\xi) nt}-1}
                    {2\, \sinh(\xi t) \sinh( (1+\xi) t) \cosh(t)}
              +\frac{n}{2\, \sinh(t\xi)} -2\, d_{s,n}\, e^{-2 t}
         \Big]\, \bigg\}\, .\espneg\label{norm}
\end{align}
In the previous formulae we use the notations,
\begin{align}
     &{\cC}_1\equiv G(-i\pi)=\exp \bigg\{\,-\int_0^{\infty}\frac{dt}{t}
               \frac{\sinh^2(t/2)\, \sinh((\xi-1) t)}
                    {\sinh(2t)\, \cosh(t)\,  \sinh(\xi t)}\, \bigg\}\, ,
           \label{C1}\\
     &{\cC}_2=\exp \bigg\{\, 4\, \int_0^{\infty}\frac{dt}{t}
               \frac{\sinh^2(t/2)\, \sinh((\xi-1) t)}
                    {\sinh(2t)\, \sinh(\xi t)}\, \bigg\}\, ,
           \label{C2}
\end{align}
and $\xi$ is given by\ \eqref{msdfh}.

\section{Numerical coefficients for equation\ \eqref{mnxchyt}}

We collect in this appendix the explicit expressions of the coefficients
$u_1$, $u_2$, $v_1$ and $v_2$ which occur in
the expansion\ \eqref{mnxchyt}. 

\begin{equation}\label{uu}
\begin{aligned}
   u_1 =\, & \frac{n^2-4s^2}{16}\,  
          \Big(\, T_s\Big(\frac{n}{2}\Big) -\frac{3}{2}\, \Big) 
         + \frac{s(s-1)}{4}\, ,
                             \\
   u_2 =\, & \frac{(n^2-4s^2)(n^2+4s^2-8)}{3072}\ 
          T_s''\Big(\frac{n}{2}\Big)
         + \frac{n(n^2-4)}{192}\, T_s'\Big(\frac{n}{2}\Big)
                             \\ 
         &\quad+ \frac{3n^2-4}{192}\, T_s\Big(\frac{n}{2}\Big)
         - \frac{s(s+2)}{192} - \frac{11 n^2}{768}
         +\frac{c}{24} +  \frac{c_2\, (n^2-4s^2)}{32}\, ,
\end{aligned}
\end{equation}
where
\begin{align*} 
  &T_s(z) = \psi(z+s) + \psi(-z+s)+ 2\gamma_E + 2c \, ,\\  
  &T_s'(z) = \partial_z\, T_s(z) \,,\qquad
   T_s''(z) = \partial_z^2\,  T_s(z) \, ,
\end{align*}
and $\psi(z)=\partial_z
\log\Gamma(z)$.
The constants $c$ and $c_2$ are the same as in Eqs.~\eqref{shbcv},
\eqref{bxvcf}.
The coefficients $v_1$ and (using the expressions $u_{1,2}$ from 
\eqref{uu}) $v_2$ are
\begin{equation}
\begin{aligned}
   v_1 =\, & \frac{n(n^2-4s^2)}{128}\, T_s'\Big(\frac{n}{2}\Big) 
           +\frac{n^2-4 s^2}{64}\,  T_s^2\Big(\frac{n}{2}\Big) 
                    \\
        &\quad - \frac{3 n^2+4 s (2-5s)}{64}\, T_s\Big(\frac{n}{2}\Big)
           +\frac{7 n^2+4 s(10-17 s)}{128}  + \frac{u_1}{2}\, ,
                    \\ 
   v_2 =\, & \frac{(n^2-4 s^2)(8-3 n^2)}{3072}\, 
                  T_s''\Big(\frac{n}{2}\Big)\\
        &\quad - \bigg(\, \frac{n(n^2-4 s^2)}{128}\, T_s\Big(\frac{n}{2}\Big)
                       + \frac{n(n^2+4s^2-8 s-8)}{ 256}\, \bigg)\, 
                       T'_s\Big(\frac{n}{2}\Big)
                    \\  
        &\quad - \frac{n^2-4 s^2}{192} \, T^3_s\Big(\frac{n}{2}\Big) 
               - \frac{n^2+4s(s-2)}{128}\, T^2_s\Big(\frac{n}{2}\Big)
               - \frac{n^2-8(s^2-s+1)}{128 }\ T_s\Big(\frac{n}{2}\Big)
                    \\  
        &\quad - \frac{n^2-4 s^2}{128}\, \big( 2\, c_2-14\, \zeta(3)-3\big) 
               - \frac{s(s-4)}{ 64} + \frac{u_1}{8} + \frac{v_1}{2} 
               + \frac{3u_2}{2} - \frac{c}{8}\ .
\end{aligned}
\end{equation}

\section{Spin-spin correlation 
functions for $\epsilon\ll 1$}\label{ap-RG}

In this appendix, 
we give  RG improved expansions for the different-time
spin-spin correlation functions which were
discussed in Section \ref{secXXX}:

\begin{multline}
  \bracket{\, T\, \sigma_{l+j}^x(t)\, 
 \sigma_{j}^x(0)\, }
  \sim\sqrt{\frac{2}{\pi^3}}\ 
   \frac{e^{-(c+\gamma_E+\frac{1}{2}\log(8\pi)+\frac{1}{4})\epsilon^2}}
        {(\sqrt{l_+l_-})^{1+\epsilon^2}\ 
         ( g^2_{\perp} )^{\frac{1-\epsilon^2}{4}}}
            \displaybreak[0] \\  
  \quad\, \times e^{u_1^x g_{\parallel}+u_2^x g_{\parallel}^3}\, \Big\{\, 1+
        g_{\perp}^2 (v_1^x -v_2^x g_{\parallel})+O(g^4)\, \Big\}
-\frac{(-1)^{l}}{\pi^2}\,
     \frac{(g^2_{\perp})^{\frac{\epsilon^2}{4}}\
           e^{-(\gamma_E+\frac{1}{2} \log(8\pi) )\epsilon^2}}
          {(\sqrt{l_+ l_-})^{2+\epsilon^2}}
            \displaybreak[0] \\
  \quad\, \times
   \exp\bigg\{\, \frac{g_{\parallel}}{2}+\frac{c}{2}\,  g_{\perp}^2
                    +\Big(\frac{1}{96}+\frac{c}{8} \Big)\,  g_{\parallel}^3
                    -\Big(\frac{1}{32}-\frac{c}{8}
                    -\frac{c^2}{2} \Big)\,  g_{\parallel}g_{\perp}^2
                    +O(g^4)\bigg\}
            \displaybreak[0] \\
\label{bcfr} 
\times \bigg[\,  \frac{1}{2}\ \Big( \frac{l_+}{l_-}+\frac{l_-}{l_+} \Big)
 +\frac{{g_\perp }}{4 }\, 
    \Big(g_{\parallel}+ \Big(  c-\frac{1}{2} \Big)\, 
 g^2_{\parallel}+c\, g_{\perp}^2+O(g^3) \Big)\,  \bigg] +\cdots
\end{multline}
and
\begin{multline}
   \bracket{\,  T\, 
\sigma_{l+j}^z(t)\,  \sigma_{j}^z(0)\, }
       \sim\sqrt{\frac{8}{\pi^3}}\ \ 
        \frac{(-1)^l\ \sqrt{g_{\perp}}\ 
e^{(\frac{1}{4}+c)\epsilon^2}}
             { \sqrt{l_+l_-}\ (g_{\parallel}+ g_{\perp})}
                    \\
    \quad\, \times
        e^{u_1^z g_{\parallel}+u_2^z g_{\parallel}^3
           +g_{\perp}^2(v_1^z -v_2^z g_{\parallel})}\ 
        \Big(1-  g_{\perp}\, (w_1 -w_2 \, g_{\parallel}
        +w_3\, g_{\parallel}^2
        +w_4\, g_{\perp}^2)+O(g^4)\, \Big)
                    \\
    \quad\, -\frac{1}{\pi^2}\ 
            \frac{1}{l_+l_-\, (1 - \frac{g_{\parallel}}{2})}\ 
        \bigg[\,  
        \frac{1}{2}\ \Big( \frac{l_+}{l_-}+\frac{l_-}{l_+} \Big)\, 
        \Big(\,  1+\Big(c-\frac{1}{4} \Big) \frac{g_{\perp}^2}{2}
                    \\
    + \Big(2 c^2+c-\frac{1}{4}\Big)\, 
              \frac{g_{\perp}^2g_{\parallel}}{4}+O(g^4)\, \Big)
           +\frac{g_{\perp}^2}{4 } 
            \Big(\, 1+ \Big(\, 2 c-\frac{1}{2}\, \Big)\, g_{\parallel}
                                      +O(g^2)\,  \Big)\, \bigg]
           +\dots\, .
    \label{bcfr1}
\end{multline}
In these expressions, the constants are given by
\begin{xalignat*}{2}
  &u_1^x=\frac{3}{8}-\frac{c}{2}\, , &
  &u_2^x=-\frac{1}{64}-\frac{\zeta(3)}{48}-\frac{c_2}{8}\, ,\\
  &v_1^x=-\frac{1}{32}+\frac{c}{8}-\frac{c^2}{4}\, , &
  &v_2^x=-\frac{5}{128}-\frac{41}{96}\, \zeta(3)+\frac{c^3}{6}
       -\frac{c_2}{8}\, ,
\end{xalignat*}
and
\begin{xalignat*}{2}
   &w_1=c\, , &
   &w_3=-\frac{1}{16}-\frac{7}{12}\, \zeta(3)+\frac{c}{8}+\frac{c^2}{2}
           +\frac{c^3}{6}+\frac{c_2}{4}\, ,\\
   &w_2=\frac{1}{8}-\frac{c(1+c)}{2}\, , &
   &w_4=-\frac{1}{32}-\frac{13}{24}\, \zeta(3)-\frac{ c}{8}
           +\frac{ c^2}{4}-\frac{c^3}{6}-\frac{c_2}{4}\, ,\\
   &u_1^z=\frac{3}{8}+\frac{c}{2}\, , &
   &u_2^z=\frac{1}{64}+\frac{\zeta(3)}{48}+\frac{c}{8}+\frac{c_2}{8}\, ,\\
   &v_1^z=-\frac{5}{32}+\frac{5c}{8}+\frac{3c^2}{4}\, , &
   &v_2^z=\frac{11}{128}+\frac{71}{96}\, \zeta(3)+\frac{c}{4}
           -\frac{5\,  c^2}{4}-\frac{2}{3} c^3+\frac{c_2}{8}\, .
\end{xalignat*}
The running couplings $g_{\parallel}, \, g_{\perp}$ in
\eqref{bcfr}, \eqref{bcfr1}\ are defined by equations
\eqref{mxncb1}\ and\ \eqref{mxncb2}\ where
\begin{equation*}
   \frac{r}{r_0}=\sqrt{2\pi\, l_+l_-}\ 
   e^{\gamma_E+c+c_2\, \epsilon^2+\ldots}\, \qquad\qquad\ {\rm and}
\qquad\qquad 
   l_{\pm}=l\pm \frac{t}{\varepsilon}\, .
\end{equation*}
Setting $l_+=l_-=l$ (equal time) and $g_{\perp}=g_{\parallel}=g, \ 
\epsilon=0$ 
(isotropic limit) in\ \eqref{bcfr}, \eqref{bcfr1},
one obtains \eqref{W(n)}.


\end{document}